%% file: paper.tex
\RequirePackage{amsmath}
\documentclass[twocolumn,table]{svjour3}          
\smartqed  
\usepackage{graphicx}
\usepackage{url}
\usepackage{hyperref}
\usepackage[
final
]{diffs}
\usepackage{xcolor}
\usepackage{glossaries}
\usepackage{amsmath}
\usepackage{comment}
\include{Acronym}

\usepackage{multirow}
\usepackage{booktabs}

\usepackage{listings}

\usepackage[T1]{fontenc}
\usepackage{textcomp}
\usepackage[multiple]{footmisc}
\definecolor{yamlblue}{RGB}{7,40,119}

\definecolor{code_indent}{HTML}{CCCCCC}
\newcommand\YAMLcolonstyle{\color{black}\footnotesize\mdseries}
\newcommand\YAMLkeystyle{\color{yamlblue}\footnotesize\bfseries}
\newcommand\YAMLvaluestyle{\color{black}\fontsize{8}{8}\mdseries}

\makeatletter

\newcommand\language@yaml{yaml}

\expandafter\expandafter\expandafter\lstdefinelanguage
\expandafter{\language@yaml}
{
  keywordstyle=\color{black},
  basicstyle=\YAMLkeystyle,                                 
  stringstyle=\YAMLvaluestyle\ttfamily,
  moredelim=**[il][\YAMLcolonstyle{:}\YAMLvaluestyle]{:},   
  morestring=[b]',
  morestring=[b]",
  literate=     {-}{{-}}1
                {*}{{*}}1
                {==}{{=\,=}}1,
  columns=fixed,
}

\lst@AddToHook{EveryLine}{\ifx\lst@language\language@yaml\YAMLkeystyle\fi}
\makeatother

\lstdefinestyle{ansible}{
  language=yaml,
  frame=single,
   rulecolor=\color{black},
  basewidth = {.45em},
  showstringspaces=true,
  breaklines=true,
  captionpos=b,
  escapechar=?
}

\usepackage{pdflscape}

\usepackage{etoolbox}

\begin{document}

\declareuser{sc}{sc}{orange}
\declareuser{fk}{fk}{red}
\declareuser{je}{je}{green}
\declareuser{fz}{fz}{magenta}
\declareuser{pm}{pm}{blue}
\declareuser{revision}{revision}{blue}

\title{
Model-Based Cloud Resource \changerevision{Provisioning}{Management} with TOSCA and OCCI
}


\author{St\'ephanie Challita \and
        Fabian Korte   \and
        Johannes Erbel \and 
        Faiez Zalila \and
        Jens Grabowski \and
        Philippe Merle
}


\institute{St\'ephanie Challita \at 
            University of Rennes 1 \& IRISA/Inria, France.\\
           \email{stephanie.challita@irisa.fr}
            \and    
            Fabian Korte \& Johannes Erbel \& Jens Grabowski \at
            University of Goettingen, Germany. \\            \email{firstname.lastame@cs.uni-goettingen.de}           
           \and
           Faiez Zalila \at
           CETIC, Belgium.\\
           \email{faiez.zalila@cetic.be}
           \and Philippe Merle \at
            Inria Lille - Nord Europe \& University of Lille, France.\\
            \email{philippe.merle@inria.fr}
              }

\date{Received: date / Accepted: date}

\maketitle

\begin{abstract}
With the advent of cloud computing, different cloud providers with heterogeneous cloud services (compute, storage, network, applications, etc.) and their related Application Programming Interfaces (APIs) have emerged. This heterogeneity complicates the implementation of an interoperable cloud system. 
Several standards have been proposed to address this challenge and provide a unified interface to cloud resources.
The Open Cloud Computing Interface (OCCI) thereby focuses on the standardization of a common API for Infrastructure-as-a-Service (IaaS) providers while the Topology and Orchestration Specification for Cloud Applications \\(TOSCA) focuses on the standardization of a template language to enable the proper definition of the topology of cloud applications and their orchestrations on top of a cloud system.
TOSCA thereby does not define how the application topologies are created on the cloud. Therefore, we analyse the conceptual similarities between the two approaches and we study how we can integrate them to obtain a complete standard-based approach to manage both Cloud Infrastructure and Cloud application layers. 
We propose an automated extensive mapping between the concepts of the two standards and we provide TOSCA Studio, a model-driven tool chain for TOSCA that conforms to OCCI.
TOSCA Studio allows to graphically design cloud applications as well as to deploy and manage them at runtime using a fully model-driven cloud orchestrator based on the two standards. Our contribution is validated by successfully transforming and deploying \changerevision{two}{three} cloud applications: WordPress,
Node Cellar
\addrevision{and Multi-Tier}.

\keywords{Cloud Computing \and Standards \and OCCI \and TOSCA \and Model-Driven Engineering \and Metamodels \and  Cloud Orchestrator \and Models@run.time}
\end{abstract}

\input{Introduction}
\input{Background}

\input{Approach}
\input{Implementation}
\input{CaseStudies}
\input{Discussion}
\input{SoA}

\input{Conclusions}

{
\par\addvspace{17pt}\small\rmfamily
\trivlist\if!Availability!\item[]\else
\item[\hskip\labelsep
{\bfseries Availability}]\fi
Readers can find TOSCA Studio including TOSCA Extension, the model-driven designer and orchestrator at: \url{https://github.com/occiware/TOSCA-Studio}.
\endtrivlist\addvspace{6pt}
}

\begin{acknowledgements}
This work is supported by the OCCIware research and development project funded by French Programme d'Investissements d'Avenir (PIA). We also thank the Simulationswissenschaftliches Zentrum Clausthal-G\"ottingen (SWZ) for financial support.
\end{acknowledgements}

\bibliographystyle{unsrt}

\bibliography{References}

\end{document}

%% file: Acronym.tex
\newacronym{OCCI}{OCCI}{\textit{Open Cloud Computing Interface}}
\newacronym{TOSCA}{TOSCA}{\textit{Topology and Orchestration Specification for Cloud Applications}}
\newacronym{OGF}{OGF}{\textit{Open Grid Forum}}
\newacronym{XSD}{XSD}{\textit{XML Schema Definition}}
\newacronym{PaaS}{PaaS}{\textit{Platform-as-a-Service}}
\newacronym{OASIS}{OASIS}{\textit{Organization for the Advancement of Structured Information Standards}}
\newacronym{FSM}{FSM}{\textit{Finite State Machine}}
\newacronym{IaaS}{IaaS}{\textit{Infrastructure-as-a-Service}}
\newacronym{IDE}{IDE}{\textit{Integrated Development Environment}}
\newacronym{CAMF}{CAMF}{\textit{Cloud Application Management Framework}}
\newacronym{API}{API}{\textit{Application Programming Interface}}
\newacronym{EMF}{EMF}{\textit{Eclipse Modeling Framework}}
\newacronym{ETL}{ETL}{\textit{Epsilon Transformation Language}}
\newacronym{OOI}{OOI}{\textit{OpenStack OCCI Interface}}
\newacronym{M2M}{M2M}{\textit{Model-to-Model transformation}}
\newacronym{MDE}{MDE}{\textit{Model-Driven Engineering}}
\newacronym{MoDMaCAO}{MoDMaCAO}{\textit{Model-Driven Configuration Management of cloud Applications with OCCI}}
\newacronym{CMS}{CMS}{\textit{Content Management System}}
\newacronym{PIM}{PIM}{\textit{Platform Independent Model}}
\newacronym{PSM}{PSM}{\textit{Platform Specific Model}}
\newacronym{VM}{VM}{\textit{Virtual Machine}}
\newacronym{OCL}{OCL}{\textit{Object Constraint Language}}
\newacronym{CAMP}{CAMP}{Cloud Application Management for Platforms}
\newacronym{EDMM}{EDMM}{Essential Deployment Metamodel}
\newacronym{UML}{UML}{Unified Modeling Language}
\newacronym{ELK}{ELK}{Elasticsearch Logstash Kibana}

%% file: Introduction.tex
\section{Introduction}
\label{sec:intro}

With the growth  of cloud computing, plenty of proprietary cloud APIs have emerged which made it hard for cloud costumers to switch between different cloud providers. To tackle the problem of this \textit{cloud provider lock-in}, consortia have been formed to develop common standards for interfacing with cloud resources. The \gls{OCCI}~\cite{occiCore}, developed by the \gls{OGF}\footnote{\url{https://www.ogf.org/ogf/doku.php/start}}, \footnote{All URLs have been last retrieved on \today.}, thereby aims to provide a standardized managing interface, enabling the customer to manage cloud resources. It has been initially published in 2010 and several open-source implementations have been developed since then supporting all major open-source cloud middleware frameworks, including OpenStack\footnote{\url{http://www.openstack.org}}, OpenNebula\footnote{\url{http://opennebula.org}} and CloudStack\footnote{\url{https://cloudstack.apache.org}}.

At a higher level of abstraction, the \gls{OASIS}\footnote{\url{https://www.oasis-open.org}} developed the \gls{TOSCA}, a template format that aims to standardize the definition of application topologies for cloud orchestration. As such, it enables the customer to define the topology of the cloud application in a reusable manner and to deploy it on TOSCA compliant \changerevision{IaaS}{} clouds. TOSCA has been initially published in 2013 and major industrial cloud providers such as IBM Cloud are supporting it~\cite{breiter2014software}. 
In contrast to OCCI, TOSCA does not define how the topologies are programmatically created on the cloud infrastructure and leaves the implementation to the cloud provider. The latter is a complex and error-prone task, it requires expertise in the technical details of the target cloud API.

While the approaches of TOSCA and OCCI are different, both define a model for cloud resources. The goal of this work is to identify the conceptual similarities and differences between the two models and provide a mapping between them where possible. Such a mapping is the first step for building a fully model-driven cloud-provider agnostic cloud orchestrator that leverages both TOSCA and OCCI for portable application and infrastructure provisioning and deployment. 

An initial mapping of the two standards was introduced in~\cite{glaser2017tosca2occi}.
In this article, we extend this mapping to support a complete coverage of both standards and we concretely implement our approach based on \gls{MDE} techniques to conceive with a high level of abstraction, verify, deploy and adapt cloud applications.
The similarities and differences between the two standards are defined via transformation rules between the concepts of their metamodels. 
These rules are outputted as a \addrevision{TOSCA} model, called \textit{TOSCA Extension}, that defines the necessary information about the characteristics and the management of cloud applications based on TOSCA. TOSCA Extension conforms to the OCCIware metamodel~\cite{merle2015precise,zalila2017occiware}
written in Ecore. In fact, the OCCIware approach~\cite{zalila2019model} proposes an enhanced metamodel for OCCI and a whole tool chain for managing cloud resources. 
We leverage \gls{MDE} in our approach since it has proven to be quite advantageous and is the mostly adopted methodology to rise in abstraction from the implementation level to the model level. It also reduces the cost of developing complex systems thanks to its ability of validation and artifacts generation.

\changerevision{Cloud models play an important role to capture the expectations of a cloud API and to a priori validate the correctness of its cloud configurations. These models are manually designed so far, which is prohibitively labor intensive, time consuming and error-prone.}{Proposing a metamodel for a cloud API plays an important role to capture the expectations of this API and to a priori validate the correctness of its cloud configurations. These metamodels are manually designed so far, which is prohibitively labor intensive, time consuming and error-prone.}
To address this issue, we propose to \changerevision{infer}{generate} a model-driven specification, i.e., TOSCA Extension, from the documentation of TOSCA written in YAML. 
This is a work of reverse engineering~\cite{rugaber2004model}, which is the process of extracting knowledge from a man-made documentation and re-producing it based on the extracted information. 
TOSCA Extension is at the base of construction of \textit{TOSCA Studio} which is a tool chain for TOSCA based on the OCCIware approach. TOSCA Studio is implemented in the form of a set of Eclipse plugins. It mainly contains a TOSCA Designer allowing users to design, edit and validate TOSCA-based cloud applications, as well as an OCCI Orchestrator allowing users to \changerevision{deploy and manage these applications}{deploy these applications on IaaS Clouds and manage them}, following a models@run.time approach~\cite{blair2009models}. TOSCA-Studio is publicly available online\footnote{\url{https://github.com/occiware/TOSCA-Studio}}.

In other words, we propose a standard-based and model-driven orchestrator for cloud applications. We extend the features introduced in~\cite{glaser2017tosca2occi} in the following ways:
\begin{itemize}
    \item we propose an automated, extensive and extensible approach for mapping TOSCA types towards OCCI types,
    \item we propose an automated, extensive and extensible approach for mapping predefined TOSCA topologies towards deployable OCCI configurations,
    \item we provide TOSCA Studio, a model-driven environment for graphically designing and verifying cloud applications using TOSCA concepts,
    \item and we provide an integrated plugin that ensures a concrete deployment and runtime management of these applications using an OCCI API.
\end{itemize}

\addrevision{Our contribution targets several audiences. It is useful for TOSCA users since we provide an additional tool for designing and deploying TOSCA topologies, and for the developers of orchestrators since we provide a technical contribution that could inspire them to build their own tool and easily map TOSCA topologies towards a uniform cloud API. Finally, our approach shows that the mapping between two cloud standards is real, which is interesting, in terms of knowledge, for researchers and educators working in this field.}

The remainder of this paper is structured as follows. First, we briefly introduce the models of TOSCA and OCCI in Section~\ref{sec:standards}. Then we provide a conceptual comparison, a mapping between the two models and preliminaries about model-driven orchestration in Section~\ref{sec:approach}. In Section~\ref{sec:implementation}, we implement our approach and provide a model-driven environment, called TOSCA Studio where the cloud user can design, verify and deploy cloud configurations. These configurations are deployed and maintained via the OCCI Orchestrator. \changerevision{Two}{Three} feasibility studies are discussed in Section~\ref{sec:casestudies}. Section~\ref{sec:discussion} presents the learned lessons from combining the two standards and providing a standard-based and model-driven environment for managing cloud applications.
We compare our contribution to related work in Section~\ref{sec:stateoftheart}. Finally, we draw our conclusions and give an outlook on future work in Section~\ref{sec:conclusions}.

 

%% file: Background.tex
\section{TOSCA and OCCI}
\label{sec:standards}

\begin{figure*}
    \centering
	\includegraphics[width=0.9\textwidth]{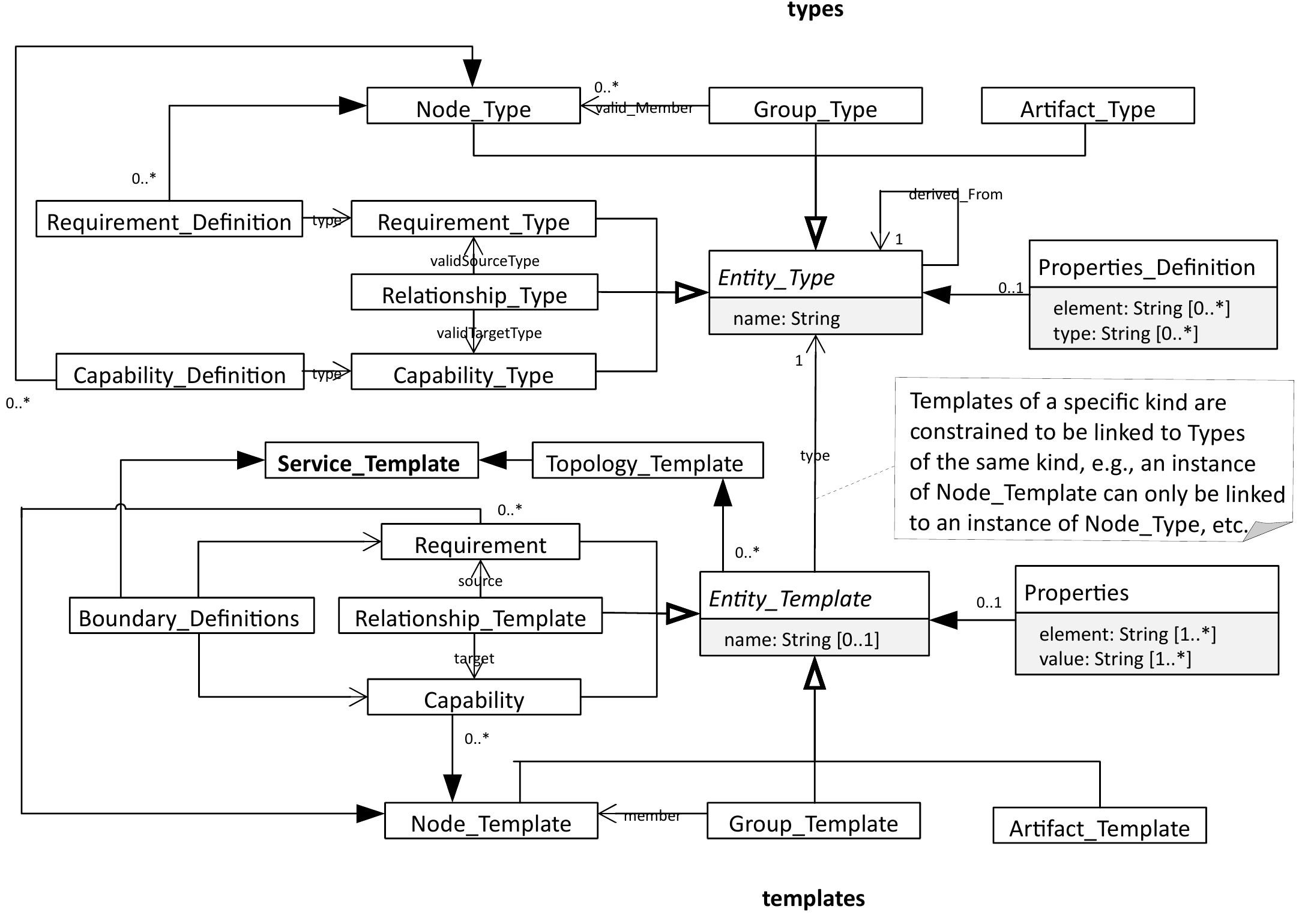}
	\caption{TOSCA metamodel.}
	\label{fig:tosca}
\end{figure*}

Both TOSCA and OCCI define languages for modeling cloud resources. Since they hence provide a model for modeling they can be seen as \textit{metamodels}~\cite{mdaguide2014}. We introduce these metamodels in the following.

\subsection{TOSCA}
\label{sec:tosca}
According to its specification~\cite{toscaSpec}, TOSCA is ``a language to describe service components and their relationships using a service topology, and it provides for describing the management procedures that create or modify services using orchestration processes''. Therefore, it is able to describe both the service structure as well as \changerevision{the processes that can be executed on this structure}{the management processes}. As the time of this writing, two versions of TOSCA exist. The first is based on XML~\cite{toscaSpec}, and the second is based on YAML~\cite{toscaSpecYAML}. While for TOSCA XML a \gls{XSD} schema exists, the TOSCA YAML version lacks of a formal metamodel. A simplified metamodel of TOSCA is depicted in Fig.~\ref{fig:tosca}.

\textit{Service\_template} captures the structure and the life cycle operations of the application. It consists of a \textit{Topology\_template} and a \textit{Plan}. Plans define how the cloud application is managed and deployed. Topology\_templa-\\tes contain \textit{Entity\_templates}, which are \textit{Node\_templates} that define e.g., the virtual machines or application components, \textit{Relationship\_templates} that encode the relationships between the Node\_templates, e.g., that a certain application component is deployed on a certain virtual machine, or \textit{Group\_templates}\footnote{Group\_templates and Group\_types are currently part of the TOSCA YAML rendering, but not part of the TOSCA XML specification.} that allow to define groups of Node\_templates, which e.g. should be scaled together. Additionally, TOSCA defines the Entity\_templates \textit{Capability} and \textit{Requirement}. Capabilities are used to define that a Node\_template has a certain ability, e.g., providing a container for running applications, and Requirements are used to define that a certain Node\_template requires a certain Capability of another Node\_template. All Entity\_templates can have \textit{Properties}, e.g., an IP address for a virtual machine, and a certain \textit{type} that references an \textit{Entity\_type}.
The Entity\_type defines the allowed Properties through \textit{Property\_definitions}, and have \textit{Interfaces}, which define the \textit{Operations} that can be executed on instances implementing the type, e.g., the termination of a certain application component, or the restart of a virtual machine. Operations have \textit{Parameters} that define their input and output. In addition to parameters for operations, TOSCA also allows to define input parameters for Plans.
\changerevision{}{Many types inherit from \textit{Entity\_type} such as \textit{Node\_type} and \textit{Relationship\_type}. The former is a reusable entity that defines the type of one or more Node Templates and the latter defines the connection between the node types. We provide more information in Section~\ref{sec:approach}.} 
Besides this abstract metamodel, the TOSCA YAML specification defines \textit{normative types} that should be supported by each TOSCA conforming cloud orchestrator.
These normative types include e.g., base types for cloud services and virtual machines. More details on the model elements can be found in \cite{toscaSpecYAML} and \cite{toscaSpec}.

\begin{figure*}
    \centering
	\includegraphics[width=0.9\textwidth]{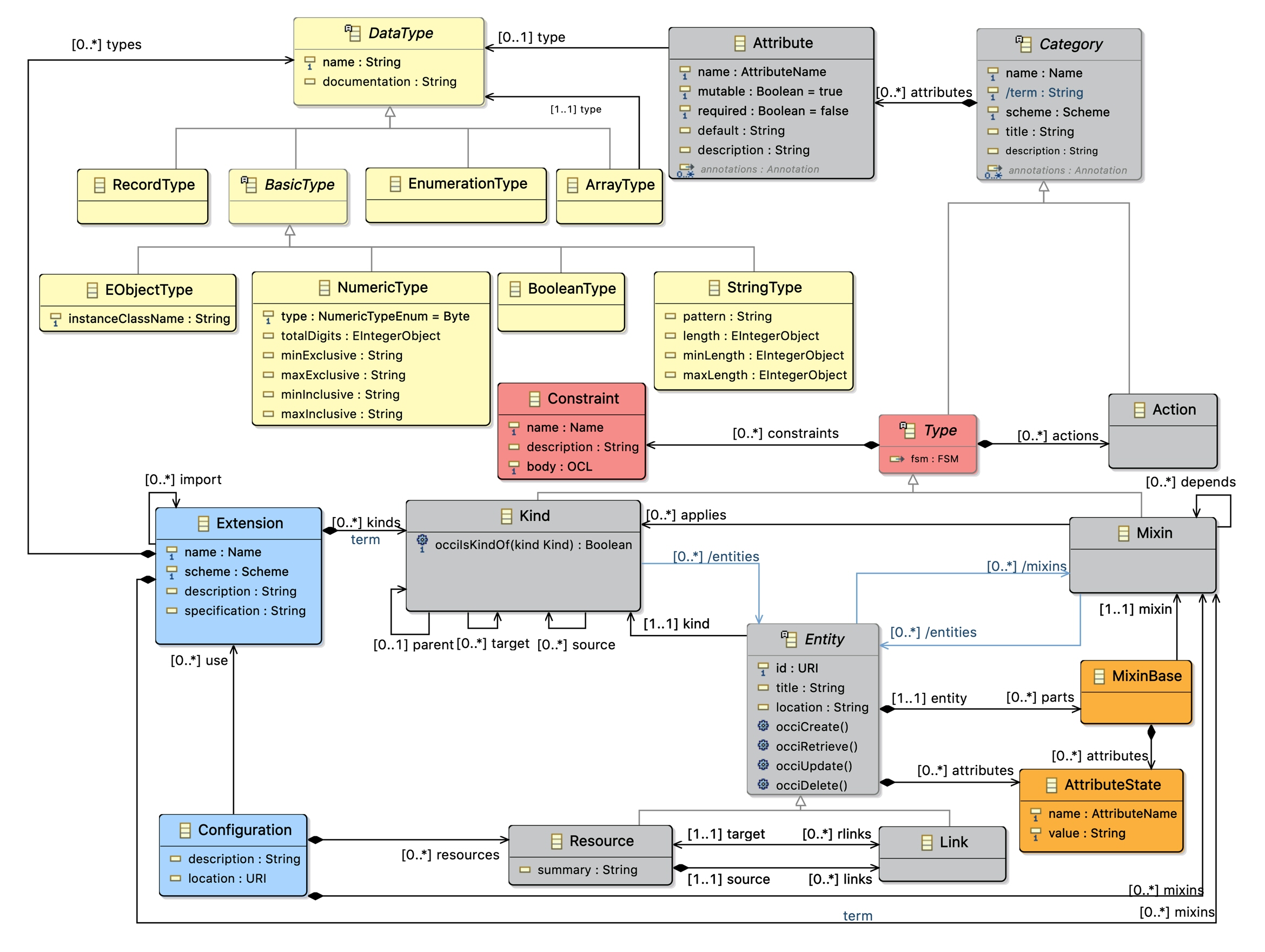}
	\caption{A subset of OCCIware metamodel (adapted from 
	\cite{zalila2017occiware}).}
	\label{fig:occiware-metamodel}
\end{figure*}

\begin{figure*}[h!]
	\centering
	\includegraphics[width=1\textwidth]{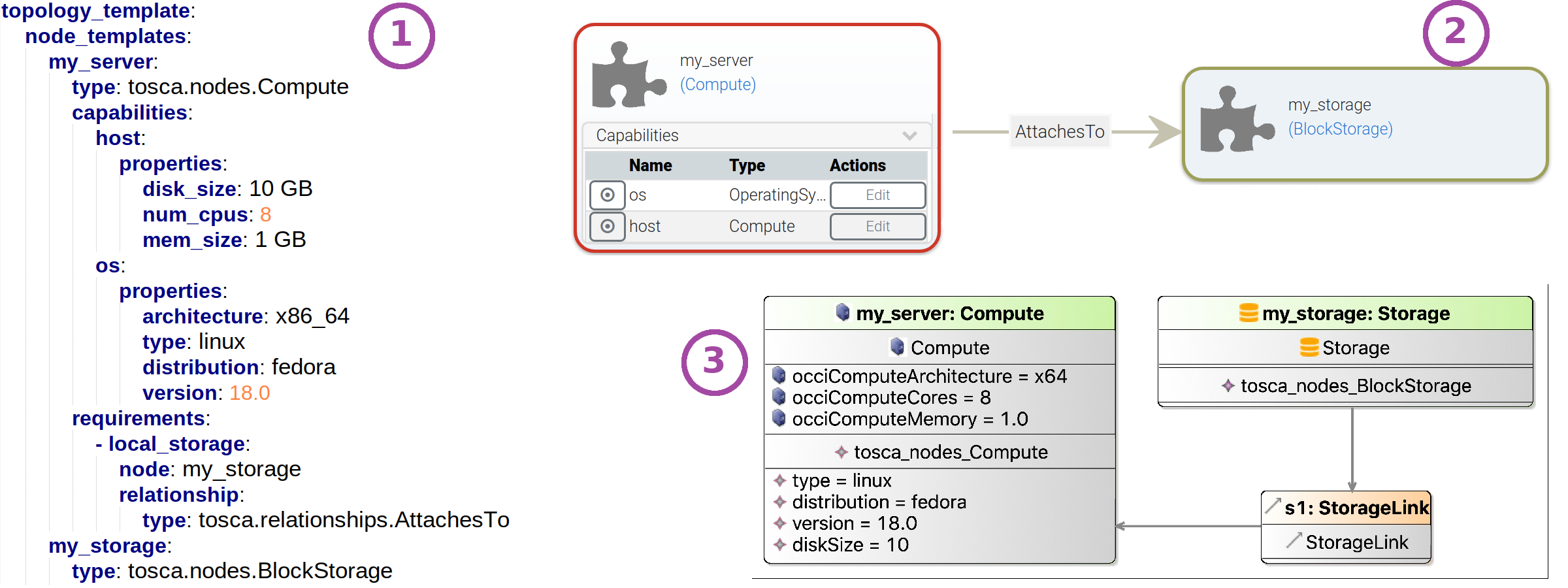}
	\caption{TOSCA vs OCCI model.}
	\label{fig:vino4tosca}
\end{figure*}

\subsection{OCCI}
\label{sec:occi}
According to the \gls{OGF}, ``OCCI is a Protocol and API for all kinds of Management tasks. It was originally initiated to create a remote management API for IaaS model based services, allowing for the development of interoperable tools for common tasks including deployment, autonomic scaling and monitoring''\footnote{\url{http://occi-wg.org}}.
The OCCI specification comprises several parts: OCCI Core model, OCCI Extensions, OCCI Renderings and OCCI Protocols.
The OCCI Core Model~\cite{occi-core-12} defines a model for cloud resources and their dependencies. In addition to the OCCI Core Model, OCCI Extensions define extensions of the core model to be used for a specific domain. Several extensions are already standardized, e.g., the OCCI Infrastructure Extension~\cite{occiInfrastructure}, which defines compute, network and storage resources for IaaS clouds, and the OCCI Platform Extension for the \gls{PaaS} domain, that defines additional resources for the Platform Service level. The extensions also define the links that can be established between the resources, for instance \textit{StorageLink}, which connects a compute resource to a storage resource and \textit{NetworkInterface} and \textit{IPNetworkInterface} which connects a compute resource to a network resource.
Finally, OCCI Renderings define how the OCCI Core Model can be interacted with, e.g., the OCCI HTTP Protocol~\cite{occiHTTPRendering} that defines how OCCI resources can be managed over the HTTP protocol. 
The OGF does not provide a formal metamodel for OCCI. 
This gap has been addressed by the OCCIware approach~\cite{zalila2017occiware,zalila2019model} and we adopt the OCCIware metamodel in the scope of this work. 
The OCCIware metamodel defines precise semantics of OCCI Core which is composed of eight elements that are represented in the grey boxes of Fig.~\ref{fig:occiware-metamodel}. The \textit{Category} is the base type for all other classes and provides the necessary identification mechanisms. Categories have \textit{Attributes} that are used to define the properties of a certain class, e.g., the IP address of a virtual machine.
Three classes are derived from Category: \textit{Kind}, \textit{Action}, and \textit{Mixin}. Kind defines the type of a cloud entity, e.g., a compute resource and Mixins define how an entity can be extended. Both have Actions that define which actions can be executed on an entity. The cloud entities themselves are modeled by the class \textit{Entity}, which provides the base class for cloud \textit{Resources}, e.g., virtual machines, and \textit{Links} that define how the resources are connected.
Moreover, the OCCIware metamodel explicitly introduces, among others, the two key concepts: \textit{Extension} and \textit{Configuration} as represented in the blue boxes of Fig.~\ref{fig:occiware-metamodel}. An OCCI \textit{Extension} represents a specific \removerevision{application} domain \addrevision{such as infrastructure, platform, security, etc.} and an OCCI \textit{Configuration} defines a running system, \addrevision{i.e., an instance model}. It represents an instantiation of one or several OCCI extensions. 
The yellow boxes of the metamodel represent the \textit{DataType} concepts that define exact types of the attributes such as \textit{StringType}, \textit{RecordType}, \textit{ArrayType}, etc.
In addition, the \textsc{OCCIware metamodel} introduces the \textit{Constraint} notion (pink box in Fig.~\ref{fig:occiware-metamodel}) allowing the cloud architect to express business constraints related to each cloud computing domain. The constraints can be imposed on OCCI Kinds and Mixins.

\addrevision{We provide Fig.~\ref{fig:vino4tosca} to better understand the correspondence between the two standards. First, we provide a simple TOSCA topology example, presented as a YAML file (1). This topology is composed of a \textit{tosca.nodes} \textit{.Compute} node named \textit{my\_server} and a\\ \textit{tosca.nodes.BlockStorage} node named \textit{my\_storage}. There is a relationship \textit{tosca.relationships.AttachesTo} that connects these two nodes. Later on, we use Winery\footnote{\url{https://winery.readthedocs.io/en/latest/user/getting-started.html}} to model this topology (2). Winery implements the visualization concept specified by Vino4TOSCA~\cite{breitenbucher2012vino4tosca}.
Eventually, we model the same topology in an OCCI dialect (3). 
One can see that the \textit{tosca.nodes.Compute} is translated into an OCCI Mixin that is applied on an OCCI \textit{Compute} resource. An OCCI Mixin can be seen as an interface that adds additional properties to a Resource, if required even at runtime. In this example, the \textit{tosca\_nodes}\\\textit{\_Compute} Mixin is applied to an OCCI Compute to add TOSCA specific properties like the \textit{type}, \textit{distribution}, \textit{version} and \textit{diskSize}.
Due to these extending capabilities of OCCI Mixins, we choose them to represent TOSCA concepts.}

%% file: Approach.tex
\section{A Standard-based and Model-driven Approach for Managing Cloud Applications}
\label{sec:approach}

In this section, we present our contribution that allows cloud application management by relying on \gls{TOSCA} and \gls{OCCI}. First, we give an overview of the proposed architecture and then we present how TOSCA concepts can be mapped to those of OCCI. We also describe how the automated generation of appropriate deployment artifacts can be achieved. 

\subsection{Overview}
\begin{figure}
	\centering
	\includegraphics[width=0.48\textwidth]{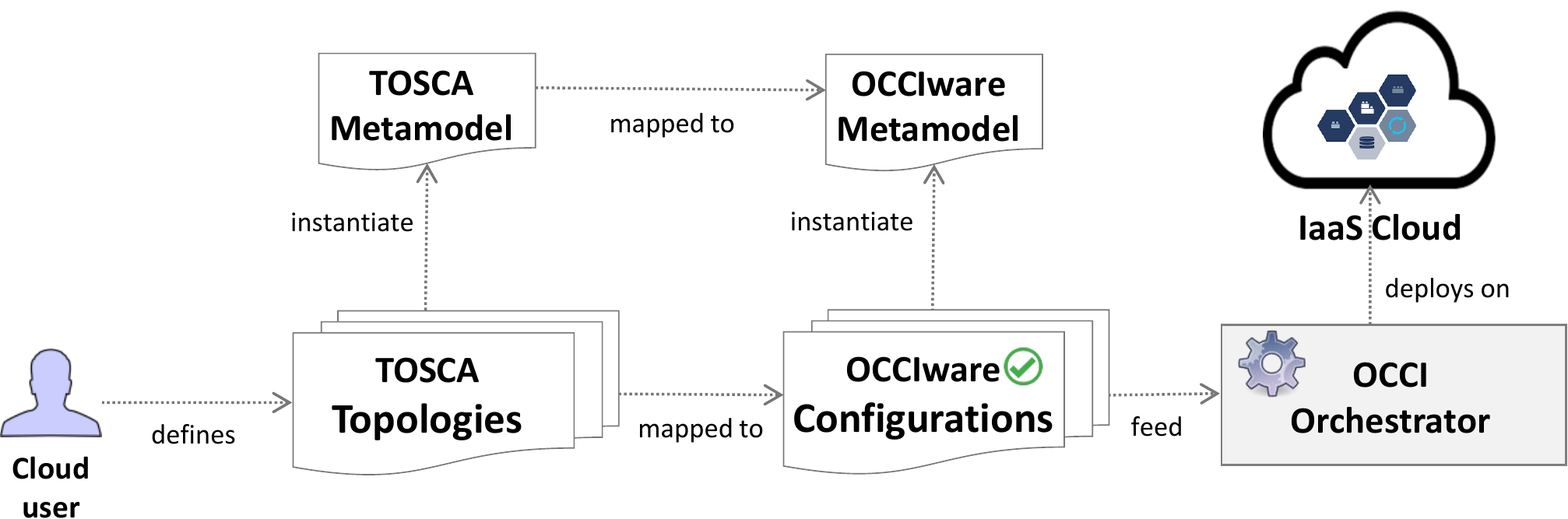}
	\caption{Approach overview.}
	\label{fig:overview}
\end{figure}

Our contribution ensures a standard-based approach to handle cloud applications in production environments.
An overview of the proposed architecture is shown in Fig.~\ref{fig:overview}.
The architecture is composed of three parts: (1) The \textsc{TOSCA Metamodel} that is mapped to the \textsc{OCCIware Metamodel}, (2) the \textsc{TOSCA Topologies} that are mapped to \textsc{OCCIware Configurations} and (3) the \textsc{OCCI Orchestrator}. 
The \textsc{TOSCA Metamodel} provides an expressive model for cloud applications by relying on an appropriate formalization of TOSCA concepts. The concepts of this model are mapped to OCCIware concepts (cf. Sections~\ref{sub:normativetypes} and ~\ref{sub:customtypes} for more details). 
The \textsc{TOSCA Topologies} describe the structure of cloud applications. They instantiate the concepts of the TOSCA metamodel. These topologies are mapped to OCCI configurations that can be designed, edited, validated and deployed as cloud applications. Further information about TOSCA topologies can be found in Section~\ref{sub:instantiation}.
Finally, \textsc{OCCI Orchestrator} provides means to generate necessary OCCI artifacts and to deploy, via appropriate OCCI requests, the generated artifacts in the executing environment. Every artifact is handled in a seamless way thanks to the homogeneity provided by modeling principles. 

To sum up, our approach uses \gls{MDE} techniques in order to design, verify and deploy cloud applications at a high level of abstraction. In fact, our \textsc{TOSCA Model} describes explicitly concerns of a cloud application that can be instantiated and deployed in an executing environment, i.e., a IaaS Cloud.

\subsection{Mapping the two standards}
\label{sub:mapping}
While both standards define a metamodel for cloud resources, their focuses are different. The focus of OCCI is to provide a standardized API and it does not define concepts to address reusability, composability, and scalability. Instances of OCCI are not meant to be stored persistently and to be reused later on as it is the goal of TOSCA.
TOSCA on the other side does not define how the defined topology is deployed by means of API calls to the cloud provider as it is done with the OCCI HTTP rendering. Hence, both approaches have their strengths and weaknesses and it is worthwhile to investigate how to integrate them.
The mapping between the two standards is possible and is done through three stages: (1) mapping of TOSCA normative types to the OCCIware metamodel, (2) mapping of TOSCA custom types to OCCIware mixins and (3) mapping of TOSCA instantiation concepts to OCCIware instantiation concepts.
This mapping is proposed after a deep reading and understanding of both TOSCA and OCCI specifications.
Each of the mapping stages will be detailed in the following subsections.

\subsubsection{\changerevision{TOSCA normative types}{TOSCA metamodel \& normative types} to OCCIware metamodel}
\label{sub:normativetypes}
We base our mapping on the TOSCA YAML specification~\cite{toscaSpec} that defines the TOSCA normative types, and on the OCCIware metamodel~\cite{zalila2019model}. We chose the TOSCA YAML specification since it has a more concise syntax, is easier to read and is widely adopted by the community comparing to the TOSCA XML specification. 
TOSCA normative types model several types of components, called nodes, that interact through relationships.
In the following, we present the main concepts of TOSCA and how they can be related to OCCI concepts. The entirety of this mapping is presented in Table~\ref{table1}.

\begin{table*}
\caption{TOSCA2OCCI mapping: metamodeling level}
\label{table1}
\begin{tabular}{ll}
    \toprule
    \textbf{TOSCA metamodel \& normative types}   & \textbf{OCCIware metamodel} \\
    \midrule
    \rowcolor{gray!25}\textbf{Entity\_type}                         & \textbf{Mixin}\\
    description                        &  description \\
    \rowcolor{gray!25}derived\_from                      & parent\\
    \midrule
    \textbf{Property \& Attribute} & \textbf{Attribute} \\ 
    \rowcolor{gray!25}default                            & default\\ 
    required &  required\\
    \rowcolor{gray!25}type    &  DataType \\
    constraints & regular expressions \\
    \rowcolor{gray!25} valid\_values                      & EnumerationType \\
    greater\_or\_equal                 &  minInclusive\\
    \rowcolor{gray!25}min\_length                        & minLength \\
    \midrule
    \textbf{Node\_type}  & \textbf{Mixin applied to Resource} \\
   \cellcolor{gray!25}nodes.BlockStorage & \cellcolor{gray!25}Mixin applied to Storage Resource \\  
    nodes.ObjectStorage & Mixin applied to Storage Resource \\
    \cellcolor{gray!25}nodes.Compute & \cellcolor{gray!25}Mixin applied to Compute Resource \\  
    nodes.SoftwareComponent & Mixin applied to Component Resource \\
    \cellcolor{gray!25}nodes.WebServer & \cellcolor{gray!25}Mixin that depends on nodes.SoftwareComponent Mixin  \\ 
    nodes.WebApplication & Mixin applied to Component Resource \\ 
    \cellcolor{gray!25}nodes.DBMS  & \cellcolor{gray!25}Mixin that depends on nodes.SoftwareComponent Mixin and on Database Mixin \\  
    nodes.Database  & Mixin applied to Component Resource \\ 
    \cellcolor{gray!25}nodes.LoadBalancer  & \cellcolor{gray!25}Mixin applied to Resource \\
    nodes.container.Runtime &  Mixin that depends on nodes.SoftwareComponent Mixin\\  
    \cellcolor{gray!25}nodes.container.Application & \cellcolor{gray!25}Mixin applied to Component Resource \\ \midrule
    \textbf{Requirement\_type}      & \textbf{OCL Constraint}\\
    \midrule
    \cellcolor{gray!25}\textbf{Relationship\_type} &    \cellcolor{gray!25}\textbf{Mixin applied to Link}\\  
    relationships.AttachesTo & Mixin applied to StorageLink Link  \\
    \cellcolor{gray!25}relationships.ConnectsTo & \cellcolor{gray!25}Mixin applied to ComponentLink Link \\
    relationships.DependsOn   & Mixin applied to ComponentLink Link\\
    \rowcolor{gray!25}relationships.HostedOn  & Mixin applied to ComponentLink Link \\
     relationships.RoutesTo  & Mixin that depends on relationships.ConnectsTo Mixin \\
     \midrule
    \textbf{Datatype\_type}  &  \textbf{RecordType} \\  
    \cellcolor{gray!25}datatypes.Credential & \cellcolor{gray!25}CredentialRecordType \\ 
    datatypes.network.NetworkInfo  & NetworkInfoRecordType\\  
    \cellcolor{gray!25}datatypes.network.PortDef  & \cellcolor{gray!25}PortDefRecordType\\  
    datatypes.network.PortInfo  & PortInfoRecordType\\  
    \cellcolor{gray!25}datatypes.network.PortSpec   & \cellcolor{gray!25}SHORT\\
    \midrule
    \textbf{Interface\_type} & \textbf{Mixin (with 0 attribute and only actions)}\\ 
    \cellcolor{gray!25}interfaces.node.lifecycle.Standard & \cellcolor{gray!25}Mixin applied to Resource \\ 
    interfaces.node.lifecycle.Standard/start() & Component/start() or Storage/online() or Compute/start()\\ 
    \cellcolor{gray!25}interfaces.node.lifecycle.Standard/stop() & \cellcolor{gray!25}Component/stop() or Storage/offline() or Compute/stop()\\   
    interfaces.relationship.Configure  & Configure Mixin applied to Link \\
    \cellcolor{gray!25}Operation   & \cellcolor{gray!25}Action \\
    \midrule
    \textbf{Capability\_type} & \textbf{Mixin applied to Resource or Link}\\
    \bottomrule
\end{tabular}    
\end{table*}

\begin{itemize}
\item \textsc{Entity\_type} is an abstract concept used to define reusable elements in TOSCA, such as \textsc{Node\_type}, \textsc{Requirement\_type}, \textsc{Relationship\_type}, \textsc{Policy\_type} and \textsc{Capability\_type}. This matches the purpose of attributes of \gls{OCCI} \textsc{Kinds} or \textsc{Mixins}. Each Entity\_type may have a \textit{description} field that provides a description of the entity and a \textit{derived\_from} field that defines the parent this new entity derives from. They match the concepts of \textit{description} and \textit{parent} in OCCI, respectively. Each Entity\_type may also have properties or attributes that define the properties that a certain entity is allowed to have. In our approach, we can map all the elements that inherit from Entity\_types, namely Node\_types and Relationship\_types to \gls{OCCI} mixins. Their properties become attributes in \gls{OCCI}. \\

\item \textsc{Property} \& \textsc{Attribute} define the properties that a certain \textsc{Entity\_type} is allowed to have. This matches the purpose of OCCI \textsc{Attributes}. A property definition should have a \textit{type}, which matches the \textit{DataType} concept in OCCI. Constraints can be applied to the attribute type, like the \textit{valid\_values} constraint that limits the property value to values declared in a list, and the \textit{greater\_or\_equal} constraint indicating the number of an attribute. For example, CPU that represents a characteristic of an entity, e.g., Compute, is \textit{greater\_or\_equal} than 2. These two constraints (\textit{valid\_values} and \textit{greater\_or} \textit{\_equal}) become an \textit{EnumerationType} declaration and a \textit{minInclusive} value in OCCI, respectively. A property may have several optional fields, for example the \textit{required} field that indicates if a property is required or not can be matched to the \textit{required} concept in OCCI.\\

\item \textsc{Node\_type} defines \changerevision{virtual machines or application components}{the possible types of building blocks for constructing a cloud application, e.g., virtual machines, network, middleware, etc.}. The node types are separately defined for reusability purpose. In fact, the defined node types can be reused when a developer or an application architect wants to define the topology of a cloud application. Further information about the instantiation of the node types are given in Section~\ref{sub:instantiation}. The \textsc{Node\_type} concept matches a \textsc{Mixin} applied to a \textsc{Kind} \textsc{Resource} in OCCI. For example,\\ \textit{tosca.nodes.BlockStorage} becomes a mixin applied to \textit{Storage} kind in OCCI. \textit{tosca.nodes.WebServer} becomes a mixin that depends on\\ \textit{tosca.nodes.SoftwareComponent} mixin. The latter is in turn applied to \textit{Component} kind in OCCI.\\

\item \textsc{Requirement\_type} defines that a certain\\ \textsc{Node\_type} requires a certain capability of another \textsc{Node\_type}. This is encoded as \gls{OCL}~\cite{warmer2003object} constraints in OCCI mixins.\\

\item \textsc{Capability\_type} extends an \textsc{Entity\_type} with a certain ability, e.g., providing an operating system for a processor or a container for a server. This concept complies to the concept of an OCCI \textsc{Mixin}. For example, \textit{tosca.capabilities.OperatingSystem} becomes a mixin that represents the operating system of a certain node. It defines information regarding of the operating system such as its \textit{type}, \textit{distribution} and \textit{version}.\\

\item \textsc{Relationship\_type} encodes the relationships between the \textsc{Node\_types}, e.g., it encodes that a specific application component is deployed on a specific virtual machine. This becomes a \textsc{Mixin} applied to a \textsc{Kind} \textsc{Link} in OCCI. For example, \textit{tosca.rel\-ationships.AttachesTo} becomes a mixin applied to \textit{StorageLink} in the OCCI Infrastructure extension.\\

%
%

\item \textsc{Data\_type} defines complex data types of TOSCA properties. This concept matches the OCCIware\\ \textsc{RecordType} concept which is used to define structures. For example, \textit{NetworkInfo} becomes a record type which contains data about the \textit{network} attribute. Each \textsc{RecordType} has at least one \textsc{RecordField} which represents a field of the record. In our example, \textit{networkid} and \textit{networkname} are record fields of the \textit{network} attribute and expect a string type value.\\

\item \textsc{Interface\_type} defines the allowed \textsc{Operations} that can be executed on \textsc{Node\_type} or on a \textsc{Relationship\_type}. This becomes a \textsc{Mixin} that contains only \textsc{Actions} in OCCI. For example,\\ \textit{tosca.interfaces.node.lifecycle.Standard} becomes a \\mixin that contains information which operations can be performed on a node type, e.g., \textit{create, configure} and \textit{delete}.

\end{itemize}


We can map all the elements that inherit from Entity\_types, namely Node\_types, Requirement\_types, Relationship\_types, \removerevision{Policy\_types}{}, Datatype\_types, Interface\_types and Capability\_types to OCCI mixins. 
However, TOSCA introduces some additional concepts, such as Group\_types, \addrevision{Policy\_types}, and Artifact\_types, that have no one-to-one correspondents in OCCI. This issue is out of the scope of this article.


\subsubsection{TOSCA custom types to OCCIware mixins}
\label{sub:customtypes}
The TOSCA specification defines basic root types called TOSCA normative types. These are default types for describing the cloud infrastructure and application. We showed in Section~\ref{sub:normativetypes} how we mapped these concepts to the OCCIware metamodel.
However, most of the application components are not part of the normative types but extend the TOSCA normative types. These are the custom types. They are defined in several YAML files that are scattered over the internet in GitHub repositories. In fact, during our study, we observed that the community around TOSCA is quite active. Several projects, such as Alien4Cloud\footnote{\url{http://alien4cloud.github.io}}, Cloudify\footnote{\url{https://cloudify.co}}, CELAR\footnote{\url{https://github.com/CELAR/c-Eclipse}}, SeaClouds\footnote{\url{http://www.seaclouds-project.eu/}}, DICER\footnote{\url{https://github.com/dice-project/DICER}} and INDIGO-DataCloud\footnote{\url{https://www.indigo-datacloud.eu}} have raised. Each project has been defining new TOSCA custom types or modifying existing types according to its need. \addrevision{In our approach, we parse these projects and we automatically map TOSCA custom types to OCCIware mixins, since custom types inherit from TOSCA normatives types. However, some of these custom types appear to be duplicated but with different names.} 
In fact, there is no centralized repository for all these types, which leads to an inconsistent use of TOSCA types across organizations. For example,\textit{tosca.nodes.Mysql}\footnote{\url{https://github.com/alien4cloud/samples/blob/master/mysql/mysql-type.yml\#L24}} and \textit{tosca.nodes.Database.\\MySQL}\footnote{\url{http://docs.oasis-open.org/tosca/TOSCA-Simple-Profile-YAML/v1.0/TOSCA-Simple-Profile-YAML-v1.0.html}}, \textit{tosca.Rsyslog}\footnote{\url{https://github.com/openstack/tosca-parser/blob/master/toscaparser/tests/data/custom_types/nested_rsyslog.yaml}} and \textit{tosca.nodes.\\SoftwareComponent.Rsyslog}\footnote{\url{https://github.com/openstack/tosca-parser/blob/master/toscaparser/tests/data/custom_types/rsyslog.yaml}} are two couples of redundant node types that are semantically equivalent but differently defined.
In our approach, we collected 30 custom TOSCA types defined in TOSCA projects and mapped them \addrevision{automatically} and exhaustively to OCCIware mixins.  \addrevision{We show in Table~\ref{table2} a subset of the TOSCA custom types automatically mapped to OCCIware mixins.}
\addrevision{In order to add new custom types, a Cloud developer has to provide a YAML file that contains the definition of his/her own types.
These types must inherit from the TOSCA normative types using the clause ``derived\_from''.
Then, within TOSCA Studio, he/she can parse the YAML file to automatically generate the corresponding OCCIware mixins that are usable by TOSCA-Studio to design and deploy cloud applications.}

By mapping TOSCA normative types, defined in the YAML specification, and the diverse added custom types, to those of OCCIware metamodel, we designed a \textsc{TOSCA Model} which conforms to the OCCIware metamodel.

\begin{table*}
\caption{Subset of TOSCA2OCCI mapping: metamodeling level}
\label{table2}
\rowcolors{2}{white}{gray!25}
\begin{tabular}{ll}
\toprule
 \textbf{TOSCA custom types}     & \textbf{OCCIware mixins} \\
 \midrule
nodes.Apache	&	Mixin that depends on nodes.WebServer Mixin\\
nodes.SoftwareComponent.Collectd	& Mixin that depends on nodes.SoftwareComponent Mixin\\
nodes.HACompute	&	Mixin that depends on nodes.Compute Mixin\\
nodes.Database.Mysql	&		Mixin that depends on nodes.Database Mixin\\
nodes.DBMS.MySQL	&		Mixin that depends on nodes.DBMS Mixin\\
nodes.Container.Application.Docker	& 	Mixin that depends on nodes.container.Application Mixin\\
nodes.SoftwareComponent.Elasticsearch	&		Mixin that depends on nodes.SoftwareComponent Mixin\\
nodes.SoftwareComponent.Logstash	&	 Mixin that depends on nodes.SoftwareComponent Mixin\\
nodes.SoftwareComponent.Kibana	&	 Mixin that depends on nodes.SoftwareComponent Mixin\\
nodes.AbstractMysql	&	Mixin that depends on nodes.Database Mixin\\
nodes.network.Network	&	Mixin applied to Network Resource \\
nodes.network.Port	&	Mixin applied to Network Resource \\
nodes.Nodejs	&	Mixin that depends on nodes.WebServer Mixin\\
nodes.WebApplication.PayPalPizzaStore	& Mixin that depends on nodes.WebApplication Mixin\\
nodes.PHP	& Mixin that depends on nodes.SoftwareComponent Mixin\\
nodes.SoftwareComponent.Rsyslog	& Mixin that depends on nodes.SoftwareComponent Mixin\\
nodes.Wordpress	&		Mixin that depends on nodes.WebApplication Mixin\\
nodes.Nodecellar & 	Mixin that depends on nodes.WebApplication Mixin \\
nodes.MongoD & Mixin that depends on nodes.DBMS Mixin \\
\bottomrule
\end{tabular}    
\end{table*}


%

\subsubsection{TOSCA instantiation concepts to OCCIware instantiation concepts}
\label{sub:instantiation}
The TOSCA specification allows the definition of a cloud application by reusing a set of nodes that are connected to other nodes using relationships.
For the definition of provisionable elements, TOSCA defines some ready-to-use topologies that represent popular cloud applications and describe their deployment. A topology is a composition of multiple nodes that may be connected through relationships. Hence, topologies use the concepts of \textsc{Entity\_templates}, namely \textsc{Node\_templates}, \textsc{Relationship\_templates} and \textsc{Group\_templates}. We explain in the following, how TOSCA topologies can be mapped to OCCIware configurations and we provide a summary in Table~\ref{table3}.
\begin{itemize}
\item A \textsc{Topology\_template} defines a reusable and \\portable representation of the structure of an application to facilitate understanding of its functional components and eliminating unnecessary details. It consists of a set of \textsc{Node\_templates} and \textsc{Relationship\_templates}. Each \textsc{Topology\_template} is mapped into a \textsc{Configuration} in OCCI.\\

\item A \textsc{Node\_template} specifies the occurrence of a node in a topology template. Each \textsc{Node\_template} refers to a \textsc{Node\_type} and instantiates the semantics of its properties, attributes, requirements, capabilities and interfaces. It gets to be transformed into a \textsc{Resource} with a \textsc{MixinBase} in an OCCI Configuration. A \textsc{MixinBase} refers to a \textsc{Mixin} and instantiates the attributes of the referenced mixin outside the owner entity in order to separate the entity attributes from the mixin ones.\\

\item A \textsc{Relationship\_template} specifies the occurrence of a relationship between nodes in a topology template. Each \textsc{Relationship\_template} refers to a \textsc{Relationship\_type} and instantiates the semantics of its properties, attributes, interfaces, etc.  It can be transformed into a \textsc{Link} with a \textsc{MixinBase} in an OCCI Configuration. \\

\item A \textsc{Group\_template} defines a group of nodes that share some semantics, e.g. an autoscaling group is a group of \glspl{VM} that would be scaled together. It can be transformed into a number of Resources and Links in OCCI.
\end{itemize}

\begin{table}
\caption{TOSCA2OCCI mapping: modeling level}
\label{table3}
\rowcolors{2}{white}{gray!25}
\begin{tabular}{ll}
\toprule
    \begin{tabular}{l} \textbf{TOSCA instantiation}\\\textbf{concepts} \end{tabular} & \textbf{OCCIware metamodel} \\
\midrule
      Topology\_template                         & Configuration\\ 
    Node\_template              & Resource with a MixinBase\\  
    Relationship\_template & Link with a MixinBase \\
\bottomrule
\end{tabular}    
\end{table}

The mapping detailed above is used to propose a fully model-driven cloud orchestrator which we define in the next subsection.




\subsection{Model-driven cloud orchestration}
\label{sub:orchestrator}
\begin{figure}[b]
	\centering
	\includegraphics[width=0.48\textwidth]{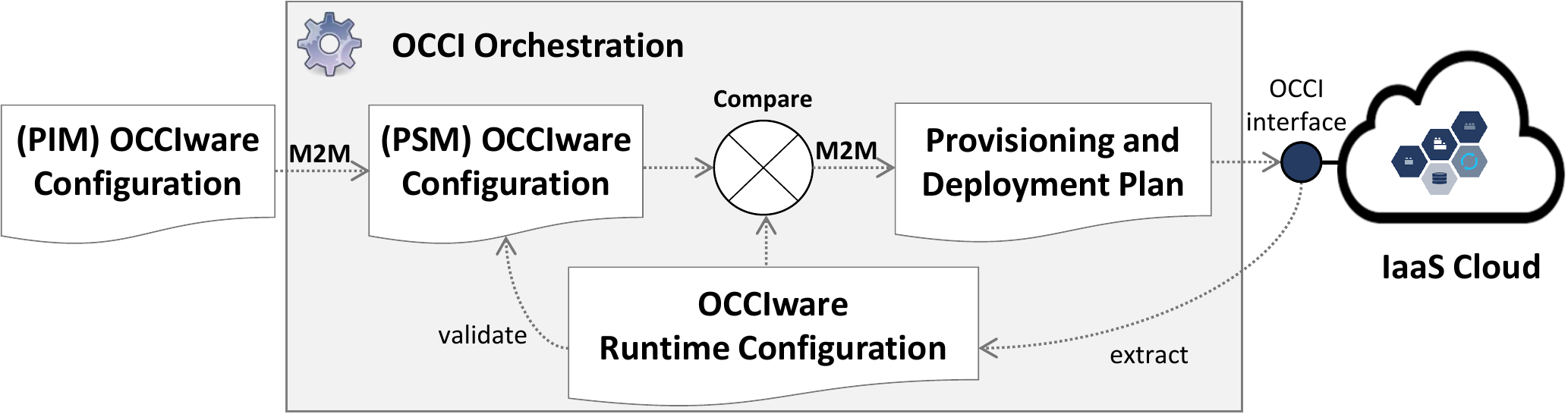}
	\caption{Model-driven cloud orchestration process.}
	\label{fig:cloudorchestration}
\end{figure}

To deploy the cloud application specified in the TOSCA topology, we use a combined approach of \gls{TOSCA} and \gls{OCCI}, as shown in Fig.~\ref{fig:overview}.
By utilizing both standards we benefit from their individual advantages while neglecting their drawbacks. Namely the matured design time capabilities of TOSCA topologies combined with the uniform interface provided by OCCI allowing to instantiate the desired cloud application \changerevision{}{using the process depicted in Fig.~\ref{fig:cloudorchestration}}. \changerevision{In the following, we describe a model-driven cloud orchestration process, depicted in Fig.~\ref{fig:cloudorchestration} that allows to adapt running cloud applications over OCCI requests. In the design phase, as described in the overview (see Fig.~\ref{fig:overview}), cloud resources are modeled according to the \gls{TOSCA} Metamodel comprising the desired infrastructure and applications to be deployed.}{} Based on the modeled TOSCA topology, an \gls{OCCI} \gls{PIM} is generated, i.e., a \textsc{(PIM) OCCIware Configuration}. The resulting configuration, containing the modeled TOSCA resources as OCCI elements, then serves as input for the \textsc{OCCI Orchestration} process initially described in~\cite{erbel18closer}. In general, the concept resembles a models@run.time approach~\cite{blair2009models} which derives imperative steps from a declarative description in order to adapt the running system. 
\changerevision{}{In the following the transformation, as well as the derivation of required OCCI requests to reach the state of the designed cloud configuration are described in more detail.}

\changerevision{}{\subsubsection{PIM to PSM Transformation}}
\changerevision{As a first step}{To allow for an automated deployment of the OCCI configuration on the target environment}, a \gls{PSM} is generated from the \textsc{(PIM) OCCIware Configuration} that contains all information required for \changerevision{an actual cloud}{the} deployment. To generate \changerevision{a}{the} \textsc{(PSM) OCCIware Configuration}, a \gls{M2M} transformation is applied that can be configured to add \changerevision{OCCI elements that are cloud provider specific.}{cloud provider specific information.} \changerevision{}{The information to be added largely depends on the elements introduced within the OCCI extensions used to handle cloud provider specifics. For example, in the extension of Paraiso et al.~\cite{paraiso2016} specialized Compute entities are designed offering attributes that can be filled to provision \glspl{VM} on individual cloud providers.}
\changerevision{Among others this comprises OCCI Operating System and Resource templates describing available sizes and images of \glspl{VM}. Moreover, this transformation adds or-\\chestration specific cloud resources required for configu-\\ration management.}{In the presented approach, the transformation is mainly used to ensure a correct functionality of the orchestration process. This comprises the addition of a mixin to each infrastructural resource which relates the OCCI id to the id assigned by the cloud provider allowing for a concrete mapping between the modeled and actual resource running in the cloud. Additionally, the transformation adds a management network to the OCCI configuration that serves two purposes. Firstly, it ensures a connection between the \textsc{OCCI interface} and each modeled \gls{VM} which is needed to deploy the modeled OCCI components via configuration management scripts. Secondly, it fulfills the requirement of specific cloud providers, such as Openstack, to declare a network to which a newly provisioned \gls{VM} gets connected. It should be noted that the transformation may be used to incorporate further provider specifics, e.g., by adding OCCI Resource and Operating system templates~\cite{occiInfrastructure} describing available sizes and images for \glspl{VM} by the target environment over OCCI terms and schemes. Finally, the resulting OCCI configuration serves as input for the orchestration process as described in the following.}

\changerevision{}{After the transformation, the orchestration process extracts the \textit{current} cloud deployment in form of an \textsc{OCCI Runtime Configuration}, and compares it to the \textit{desired} cloud deployment. Based on this comparison a \textsc{\gls{M2M}} is triggered that generates a \textsc{Provisioning and Deployment Plan} (see e.g., ~\cite{erbel18closer}, \cite{breitenbucher2014declarative} or
\cite{lushpenko2015adaptation} for automatic deployment and
provisioning workflow generation). Within this plan the OCCI requests required to manage cloud resources are sequenced in order to bring the cloud resources into the desired state. It should be noted that the \gls{OCCI} requests are generated from the elements contained within the \textsc{(PSM) OCCIware Configuration}. Finally, the provisioned and deployed OCCI model can be again extracted and synchronized to validate whether the design time concepts and constraints, specified in the TOSCA model, are met. We exemplify the proposed orchestration process in Section~\ref{sec:casestudies}.}
\subsubsection{\changerevision{}{Orchestration Process}}
\changerevision{}{To derive required adaptive steps, the current and desired cloud deployment need to be investigated. As a first step the orchestration process \textsc{extracts} the current cloud deployment in form of an \textsc{OCCI Runtime Configuration}, and compares it to the desired one. During the \textsc{comparison} each individual resource of the new and actual deployment are matched to each other based on their identity, i.e. their id and kind. Based on the match, each entity is identified as already existing, to be deleted, updated or added resulting in the corresponding instructions that need to be send to the OCCI interface. While entities marked as to be updated get their values adjusted in the runtime model and resources marked as to be deleted get undeployed and deprovisioned, the resources to be added need to be provisioned in the correct order. \\
To sequence the provisioning requests, a \textsc{\gls{M2M}} is performed on the input model generating a \textsc{Provisioning and Deployment Plan}. As a first step within this transformation, a \textit{provisioning order graph}~\cite{BreitenbCombining} is generated. This graph describes the dependencies between modeled cloud resources based on the kind of links between them. Thereafter, we remove the updated and existing resources from the graph, as they are already present in the currently deployed runtime model~\cite{erbel18closer}. As a final step, this graph is transformed into an UML activity diagram containing a sequence of provisioning actions to follow. Hereby, each action in the diagram refers to an OCCI entity from which the requests are formed. The resulting activity diagram is interpreted by the orchestration process which is responsible for sending the provisioning requests to the \textsc{OCCI interface}. After the activity diagram has been interpreted successfully, the runtime model has reached the new desired state with the infrastructural resources being started. Then the deploy, configure and start action is triggered on each application within the runtime model resulting in the deployment of the individual application components. The provisioned and deployed OCCI model can be subsequently extracted and synchronized to validate whether the design time concepts and constraints, specified in the TOSCA model, are met. We exemplify the proposed orchestration process in Section~\ref{sec:casestudies}.}

%% file: Implementation.tex
\section{Implementation: TOSCA Studio}
\label{sec:implementation}
Our approach for managing cloud applications by relying on TOSCA topologies and the OCCIware framework and API is implemented through \textbf{TOSCA Studio}.
TOSCA Studio is a model-driven tool chain for modeling and deploying cloud application topologies encoded by the TOSCA standard based on the OCCIware approach. It relies on a metamodel, called \textbf{TOSCA Extension}, defining the static semantics for the TOSCA standard in Ecore and \gls{OCL}~\cite{warmer2003object} and conforming to the OCCIware Metamodel. More specifically, TOSCA Studio is implemented as a set of Eclipse plugins that are publicly available\footnote{\url{https://github.com/occiware/TOSCA-Studio}}. It mainly contains a \textbf{TOSCA Designer} that provides users facilities for designing, editing, validating TOSCA-based cloud applications, and an \textbf{OCCI Orchestrator} that allows users to deploy and manage these applications. In this section, we detail each of our solution main components: \textbf{TOSCA Extension}, \textbf{TOSCA Designer}, and \textbf{OCCI Orchestrator}. 

\subsection{TOSCA Extension}
\begin{figure}
\centering
\includegraphics[width=0.35\textwidth]{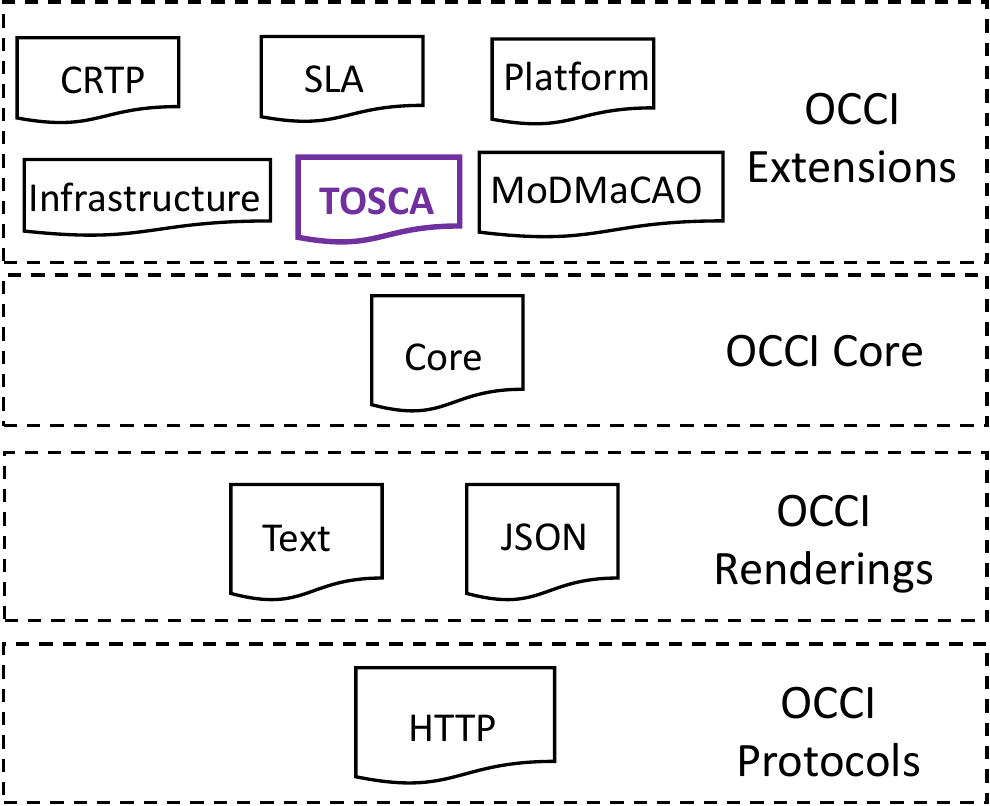}
\caption{OCCI specifications.}
\label{fig:specifications}
\end{figure}

\begin{figure}
\centering
\includegraphics[width=0.48\textwidth]{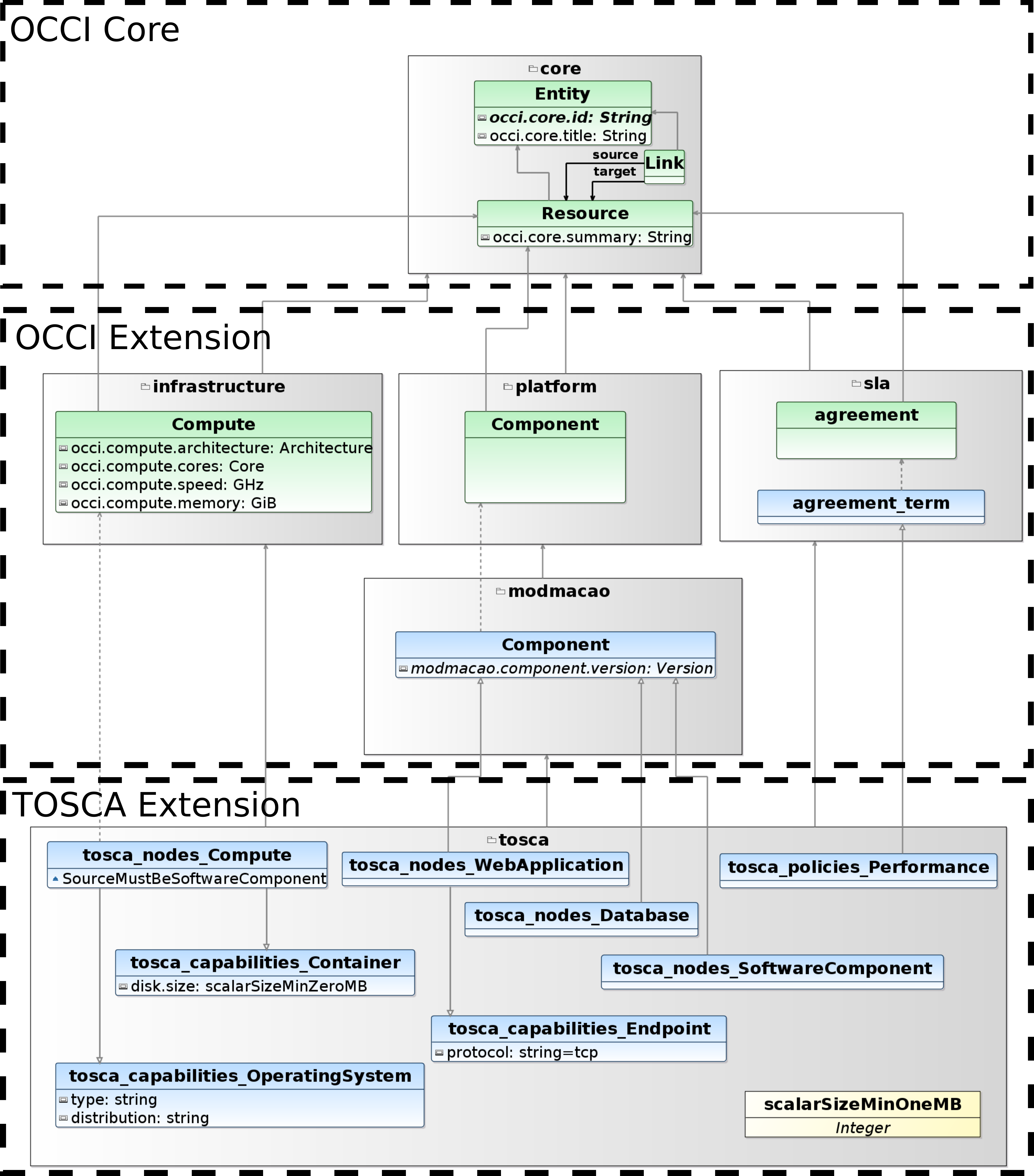}
\caption{A subset of TOSCA Extension.}
\label{fig:tosca-diagram}
\end{figure}

The mapping between the original TOSCA metamodel (in YAML) and OCCIware metamodel, detailed in Section~\ref{sub:mapping}, is encoded as an OCCI extension for TOSCA (called the \textbf{TOSCA Extension}), as depicted in the purple box in Fig.~\ref{fig:specifications}. TOSCA Extension captures all the necessary information related to the characteristics and management of TOSCA-based cloud applications. TOSCA Extension is represented in the form of a diagram in Fig.~\ref{fig:tosca-diagram}. This diagram was designed with \textsc{OCCIware Studio}, our open source model-driven development environment dedicated to OCCI~\cite{zalila2017occiware,zalila2019model}. TOSCA Extension is conceptually divided into three levels.

The top level represents the OCCI Core model, encoded with the \gls{EMF}. The middle level contains OCCI standardized extensions, which are OCCI Infrastructure, OCCI Platform and OCCI SLA. It also contains \gls{MoDMaCAO}~\cite{korte2018model}, which is an enhanced version of the standardized OCCI Platform extension.
The bottom level represents TOSCA concepts, which extend the Infrastructure, MoDMaCAO and SLA extensions, in the form of mixins.
TOSCA Extension is quite rich in concepts. It contains 68 mixins, 10 of which extend the Infrastructure extension, 33 extend the MoDMaCAO extension and 4 extend the SLA extension. The remaining concepts extend the generic Resource and Link types from the OCCI Core model. 
Fig.~\ref{fig:tosca-diagram} illustrates how TOSCA mixins extend already existing OCCI extensions, and shows the graphical output of a subset of the TOSCA Extension. For example, \textit{tosca\_nodes\_Compute} extends OCCI Infrastructure. It contains an OCL constraint \textsc{SourceMustBeSoftwareComponent} which enforces that the \textit{Compute} instance cannot run if it is not linked to a \textit{SoftwareComponent} instance. It also depends on three other mixins \textit{tosca\_capabilities\_Container}, \textit{tosca\_capabilities\_OperatingSystem} and\\ \textit{tosca\_capabilities\_Endpoint}, and therefore inherits their attributes.
TOSCA Extension also defines exact types thanks to the DataType system provided by the OCCIware metamodel. The  \gls{EMF}  validator  then checks the type constraints that are  attached to the attribute. For example, \textsc{scalarSizeMinOneMB} is translated into a \textsc{NumericType}, especially an \textsc{Integer,} containing the following constraint: \textit{minInclusive = 1}.  

To implement TOSCA Extension, we implemented a YAML parser in Java, using yamlbeans\footnote{\url{https://github.com/EsotericSoftware/yamlbeans}} library. In fact, we provide an algorithm that infers TOSCA Extension from YAML specifications. This automated extraction allows a better modeling of TOSCA concepts. So far, existing models are manually designed, which is prohibitively labor intensive, time consuming and error-prone. To address this issue, we propose a novel approach to automatically infer model-driven specification from YAML specification files of TOSCA standard.
First, using the OCCI API, we create a model, i.e., an OCCI extension. Then, the algorithm loads the content of the YAML specification files using yamlbeans library.
The types in these YAML file are grouped semantically: nodes, relationships, capabilities, data, etc.
The algorithm runs through each of this group, then for each TOSCA type it builds the corresponding OCCI mixin.
For each type, the algorithm matches its information (\textit{derived\_from}, \textit{description}, \textit{attributes}, \textit{requirements}, etc.) to corresponding OCCI concepts.
If an attribute requires a type that has not been defined yet, the algorithm keeps the name of this type in memory. When defined, the latter will be assigned to this attribute.
Eventually, we add the new mixin to the model under construction and so on. This model represents our TOSCA Extension.

Readers can find the parser code as well as our precise TOSCA Extension on GitHub\footnote{\url{https://github.com/occiware/TOSCA-Studio}}.

\begin{figure*}
\centering
\includegraphics[width=0.95\linewidth]{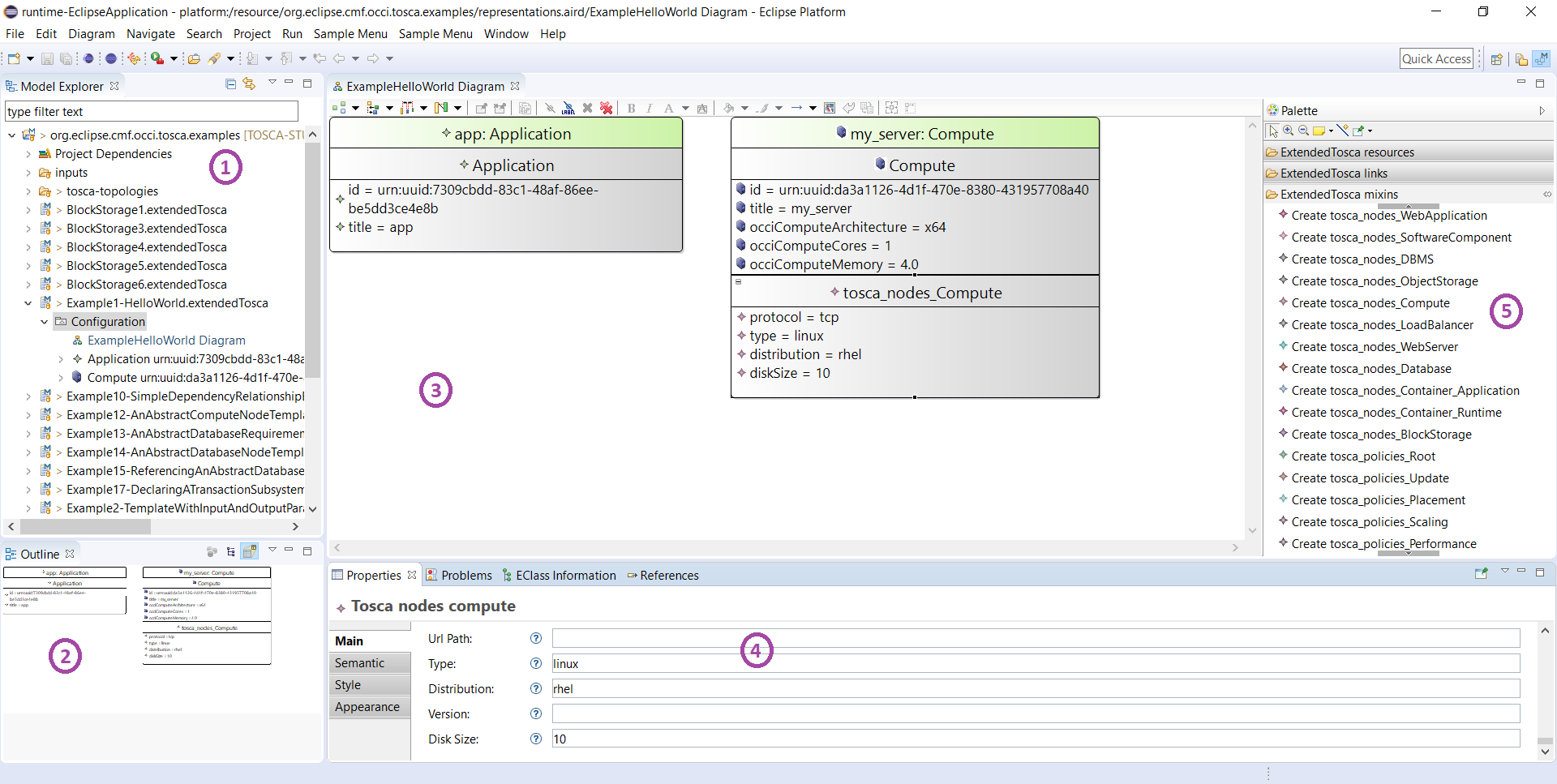}
\caption{TOSCA Designer.}
\label{fig:designer}
\end{figure*}

\subsection{TOSCA Designer}
Our approach allows cloud architects to visualize, edit and verify configured instances of cloud applications using TOSCA types defined in TOSCA Extension. To do so, we provide \textbf{TOSCA Designer}, which is a specific graphical modeler of both OCCI extensions and configurations for TOSCA. This tool is implemented on top of the Eclipse Sirius framework. 
A screenshot of our TOSCA Designer is depicted in Fig.~\ref{fig:designer}. 
Frame (1) shows the Eclipse Model Explorer used to navigate through a TOSCA project containing a TOSCA Model.
Frame (2) gives a perspective or a global view of the modeled topologies.
Frame (3) contains the topology elements. 
Frame (4) contains the Eclipse properties editor for visualizing and modifying attributes of a selected modeling element. All TOSCA elements displayed in Frame (3) can be set through their properties. 
Frame (5) displays the configuration pallet that represents the TOSCA types (normative and custom) such as: \textit{tosca\_nodes\_WebApplication, tosca\_nodes\_\\SoftwareComponent, tosca\_nodes\_WebServer, tosca\_\\nodes\_Apache}, etc.

This tool can be used in two ways:
\begin{enumerate}
\item It can take as input any existing TOSCA topology, translate it into an OCCI configuration for TOSCA and graphically represent it. To do so, we implemented a \textbf{config-generator}, which parses any existing TOSCA topology\_template written in YAML and transforms it into an OCCI configuration that conforms to the TOSCA Extension. More specifically, OCCI configuration instantiates the normative and custom types defined in TOSCA Extension. 

\item It can design cloud applications from scratch using TOSCA types from the palette of TOSCA Designer.
\end{enumerate}

Finally, TOSCA Designer checks the validity of cloud application configurations by checking all the constraints defined by used TOSCA mixins. If a constraint is false, the cloud architect must correct its cloud application configuration. When all the constraints are true, the TOSCA-based cloud application can be deployed using OCCI Orchestrator.

\subsection{OCCI Orchestrator}
To provision and deploy the transformed TOSCA application over an OCCI interface, either the required OCCI requests can be generated using TOSCA Studio and manually sequenced or the presented OCCI Orchestration process can be used. 
An implementation of the orchestration process is publicly available~\footnote{\url{https://gitlab.gwdg.de/rwm/de.ugoe.cs.rwm.docci}}.
While the presented \changerevision{concept}{orchestration process} can be generally applied on any kind of OCCI API, such as the \gls{OOI} used in~\cite{erbel18closer}, the implementation got enhanced to focus on the OCCI API provided by the \textit{OCCIware Runtime}\footnote{\url{https://github.com/occiware/MartServer}}. The OCCIware Runtime is a server that maintains a runtime model of the currently deployed cloud system which is utilized by the orchestrator. Moreover, the OCCIware Runtime server follows a plugin-based architecture for OCCI extensions modeled with OCCIware, such as the TOSCA extension. Based on these extensions, connector skeletons can be generated \changerevision{that translate incoming OCCI requests, e.g, to the API of the desired cloud provider. Thus, any kind of cloud provider can be used to provision, deploy and maintain cloud applications using the presented approach.}{that can be filled to interpret incoming OCCI requests. Among others, this mechanism is used to translate incoming OCCI infrastructure requests to the proprietary interface of the cloud provider. While our implementation provides a connector to an OpenStack Cloud, connector skeletons to further cloud providers can be easily modeled and generated within TOSCA Studio. Moreover, to address multi-cloud deployments, extensions such as the one presented in~\cite{paraiso2016} can be created to model on which specific cloud provider a VM should be provisioned including required attributes.} 

To perform the deployment of modeled components, the MoDMaCAO framework~\cite{korte2018model} is used, providing a connector which implements lifecycle operations for OCCI Applications and Components. These operations trigger the execution of configuration management scripts to deploy applications on top of \glspl{VM} as specified within the generated OCCI configuration.
\changerevision{}{Therefore, the framework provides a generic component mixin that can be extended. Each specialized component mixin is linked to the configuration management script to be executed which is deposited on the OCCIWare Runtime server making them reusable for multiple cloud deployments. In addition to typical configuration management features, the framework allows to use modeled attributes as well as runtime information, e.g., attributes of linked components and machines, within the configuration management scripts. It should be noted that the specialized component mixins, such as the ones shown in Table~\ref{table2} are automatically generated during the TOSCA to OCCI transformation.}
To perform the tasks specified within the configuration management scripts, the OCCIware Runtime server needs to be connected to the \glspl{VM} to be configured which is ensured by the \gls{PIM} to \gls{PSM} transformation. In general, for the application deployment process the orchestrator only sends a request to the OCCI API triggering the start action of the applications to be deployed. The execution of the management scripts to deploy the modeled application is then handled by the \gls{MoDMaCAO} framework. To ensure that all infrastructure requirements of the cloud topology to be deployed are met, the application deployment is only performed when the provisioned infrastructure, reflected in the runtime model, conforms to the designed state of the OCCIware configuration to be deployed.

Within TOSCA-Studio, the transformation, as well as the deployment process can be directly enacted on top of modeled or generated OCCI Configurations which allows to easily model, manage and deploy cloud resources.

%% file: CaseStudies.tex
\section{Case Studies}
\label{sec:casestudies}

To evaluate the proposed approach, we selected \changerevision{two}{three} case studies that represent popular distributed cloud applications:  a WordPress application\footnote{\url{https://fr.wordpress.com}}, a Node Cellar application\footnote{\url{http://nodecellar.coenraets.org}} and a Multi-Tier application with \gls{ELK} stack\footnote{\url{https://www.elastic.co/fr/elastic-stack}}. 
\addrevision{We chose the first two case studies because they are medium size which proves that our approach is able to handle real applications and allow us to present them in a clear manner. Moreover, these two case studies are widely used in the community. WordPress is one of the most adopted \gls{CMS} in industry and it is used as an example in TOSCA official specification~\cite{toscaSpec}, and NodeCellar is a prominent and interactive LAMP stack application that is used in Alien4Cloud and Cloudify projects. In order to prove that our approach can support larger systems, we chose the Multi-Tier use case which represents a complex system composed of \gls{ELK} stack together with a NodeCellar application.}
We relied on existing TOSCA YAML topologies for WordPress\footnote{\url{https://github.com/alien4cloud/samples/tree/master/topology-wordpress}} and Node Cellar\footnote{\url{https://github.com/alien4cloud/samples/tree/master/topology-nodecellar}}. \addrevision{For our third use case, we adapted the Multi-Tier example presented in TOSCA official specification~\cite{toscaSpec}. We only replaced the PaypalPizzaStore web application by the NodeCellar application since the former is not available anymore.} We demonstrate how our approach can design, validate and deploy these applications, and we provide a configuration model for each.

To provision and deploy the modeled cloud configurations, we used the presented model-driven cloud orchestration process. Hereby, we connected the OCCI Orchestrator to a private Openstack cloud to provision modeled infrastructure resources. To perform the deployment of the individual components, the \gls{MoDMaCAO} framework is used. The latter executes scripts of Ansible\footnote{\url{https://www.ansible.com/}},\changerevision{ a configuration management tool, to deploy, configure, and start the modeled application components.}{for which we utilized and updated already existing deployment artifacts used in the TOSCA topologies to deploy, configure, and start modeled application components.}

\subsection{WordPress}
\begin{figure*}[!ht]
\centering
\includegraphics[width=0.95\linewidth]{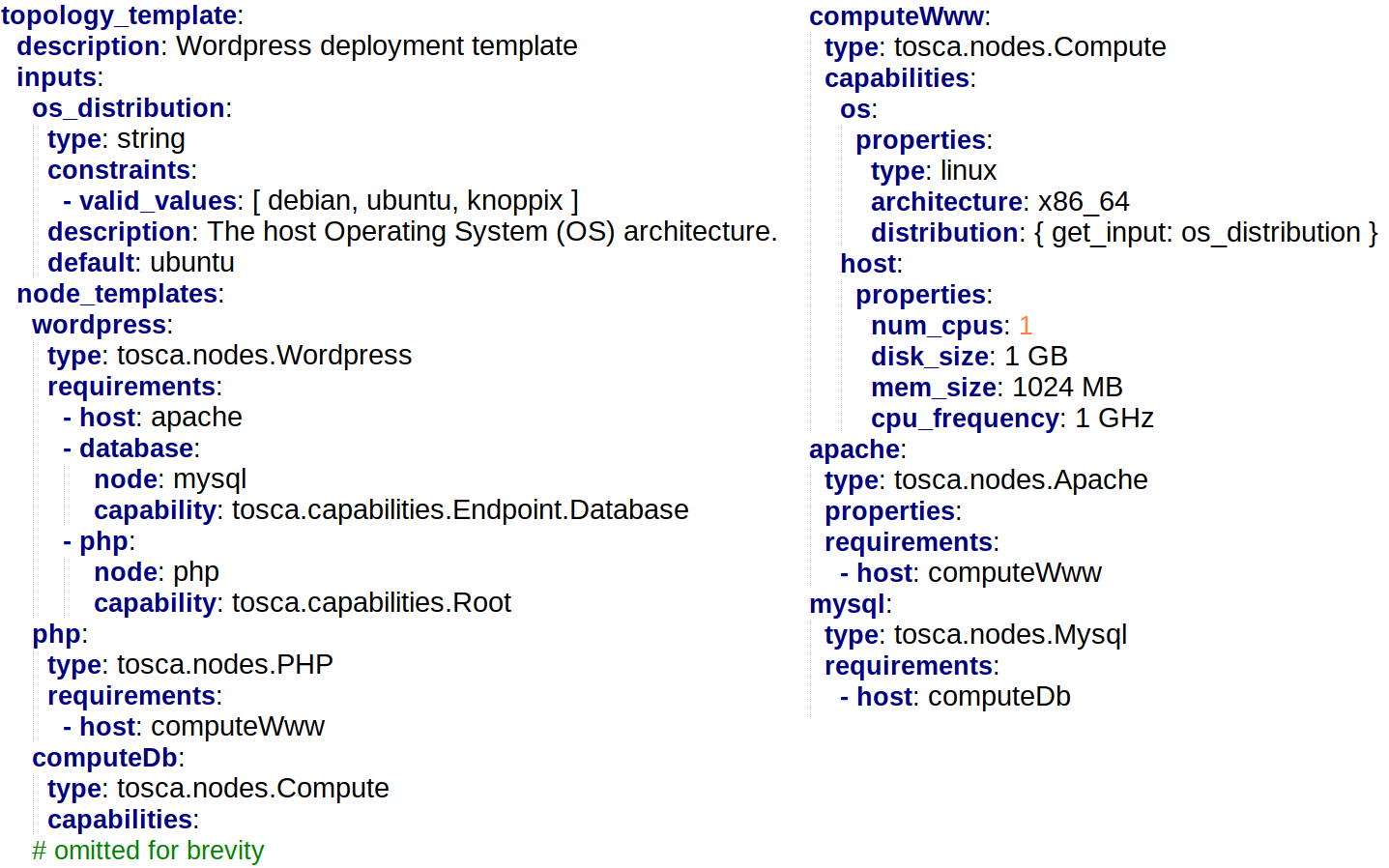}
\caption{WordPress YAML Topology.}
\label{fig:wordpress-yaml}
\end{figure*}

WordPress is an open source \gls{CMS} that allows to build custom web applications based on Apache as the web server, MySQL  as  the  relational database management system and PHP as the object-oriented scripting language. TOSCA allows to define such types, and therefore it allows to define a WordPress application. A screenshot of the WordPress topology as defined in TOSCA YAML file is depicted in Fig.~\ref{fig:wordpress-yaml}.


To validate our approach, the config-generator reuses the mixin types defined by the TOSCA Extension to be able to model a WordPress system. It is done with the help of the following mixins:
\begin{itemize}
    \item The \textbf{tosca.nodes.WordPress} mixin type abstracts the notion of a WordPress \gls{CMS} and depends on the tosca.nodes.WebApplication mixin type, which depends on the MoDMaCAO Component mixin type. It is hosted on an Apache WebServer and connected to a MySQL database and a PHP SoftwareComponent.
    \item The {\textbf{tosca.nodes.Apache}} mixin type abstracts the notion of an Apache server. It depends on the tosca.\\nodes.WebServer mixin type, which depends on the tosca.nodes.\\SoftwareComponent mixin type and therefore also on the MoDMaCAO Component mixin type.
    \item The {\textbf{tosca.nodes.Mysql}} mixin type abstracts the notion of a Mysql database. It depends on the tosca.\\nodes.Database mixin type, which depends on the MoDMaCAO Component mixin type.
    \item The {\textbf{tosca.nodes.PHP}} mixin type abstracts the notion of PHP scripting language used to develop a WordPress application. It depends on the tosca.nodes.\\SoftwareComponent mixin type, which depends on the MoDMaCAO Component mixin type.
    \item The {\textbf{tosca.nodes.Compute}} mixin type abstracts the notion of real or abstract processors of software applications such as \glspl{VM}. It is applied on the Compute resource type.
\end{itemize}

\begin{figure}
\centering
\includegraphics[width=0.48\textwidth]{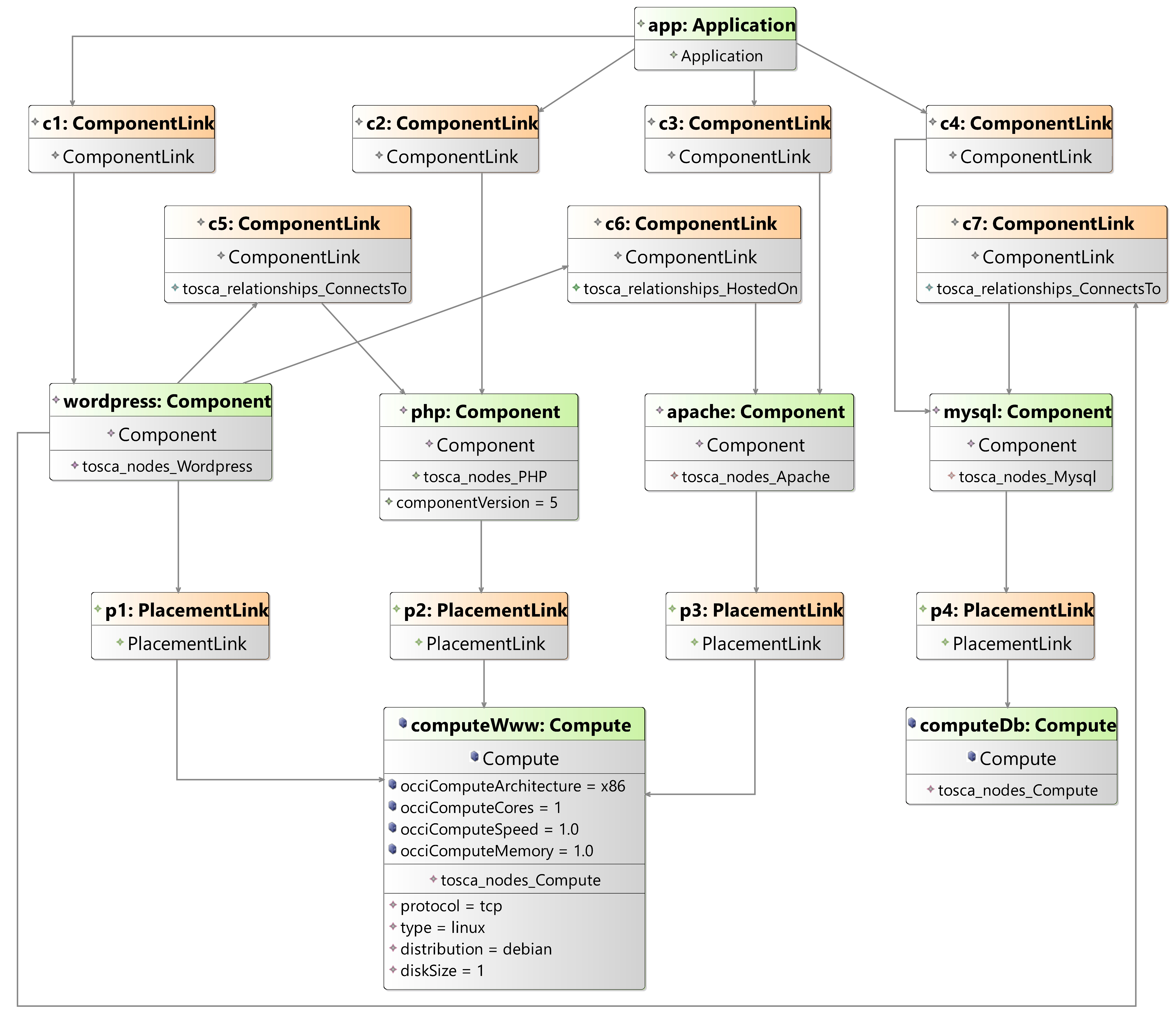}
\caption{WordPress Configuration.}
\label{fig:wordpress}
\end{figure}

Fig.~\ref{fig:wordpress} shows the model of a WordPress application that corresponds to the topology in Fig.~\ref{fig:wordpress-yaml}. It is composed of four components (\textsc{wordpress, php, apache} and \textsc{mysql})  deployed on two \glspl{VM}\\ (\textsc{ComputeWww} and \textsc{ComputeDb}).  OCCI resources and links are represented by boxes in green and orange color, respectively. The application resource  is  connected  to  the  four Component resources  via \textsc{ComponentLinks} (\textsc{c1} to \textsc{c4}). The WordPress component is connected to the PhP and MySQL components via \textsc{ConnectsTo} links (\textsc{c5} and \textsc{c7}). The WordPress is hosted on the \textsc{Apache} component via a \textsc{HostedOn} link (\textsc{c6}). Each component is placed on one \gls{VM} via a \textsc{PlacementLink} (\textsc{p1} to \textsc{p4}). Finally, the properties of all the components and \glspl{VM} are configured. For  the  sake  of  brevity,  we  omit  the depiction of \textsc{Attributes} of the components in this model. For illustration, we only keep the attributes of \textsc{ComputeWww}. We can notice that its properties declared in the YAML file of Fig.~\ref{fig:wordpress-yaml}, i.e.,  the architecture, the number of cores, the speed, the memory, the protocol, the type, the distribution and the disk size, have been correctly automatically set in the model. 

\subsection{Node Cellar}
\label{sec:nodecellar}

\begin{figure*}
\centering
\includegraphics[width=0.95\linewidth]{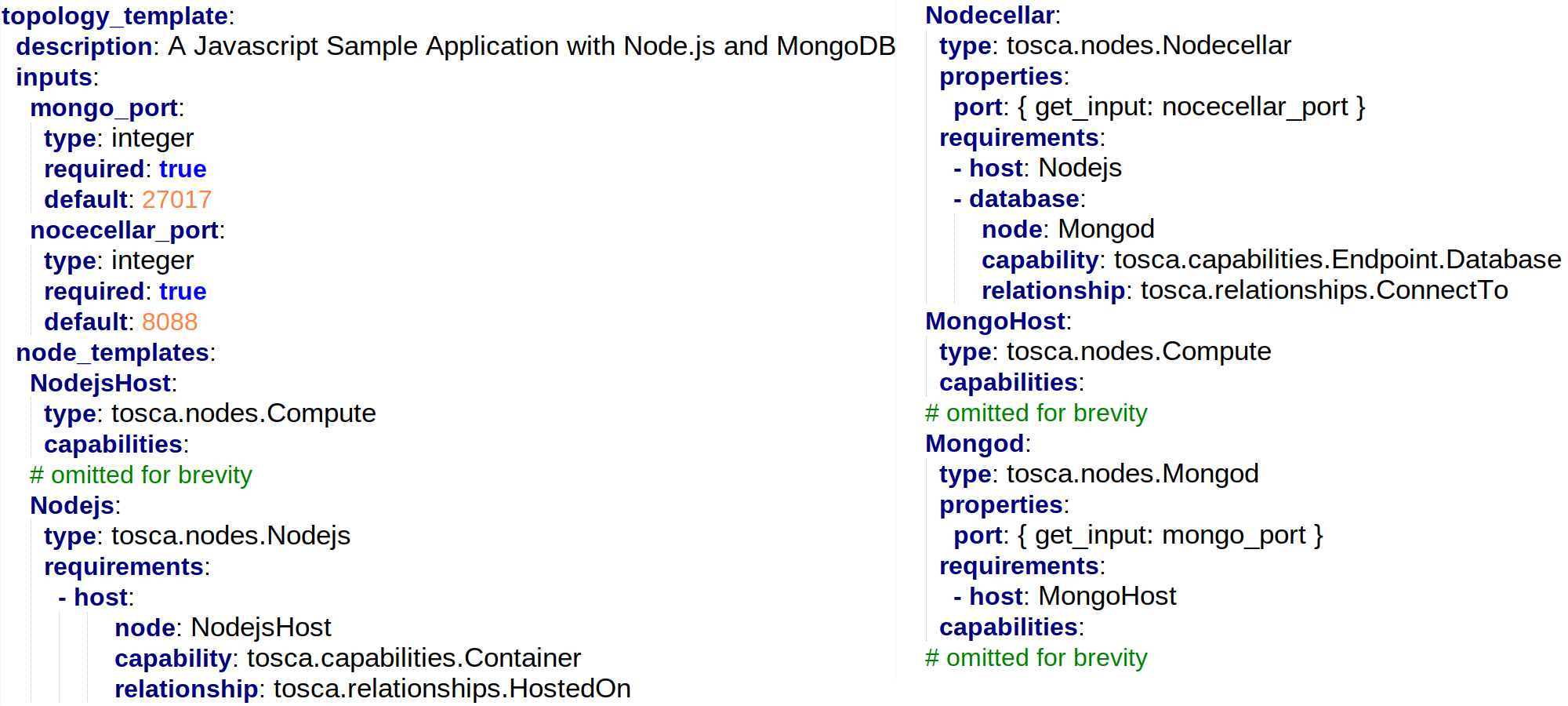}
\caption{Node Cellar YAML Topology.}
\label{fig:nodecellar-yaml}
\end{figure*}

\begin{figure}
\centering
\includegraphics[width=0.95\linewidth]{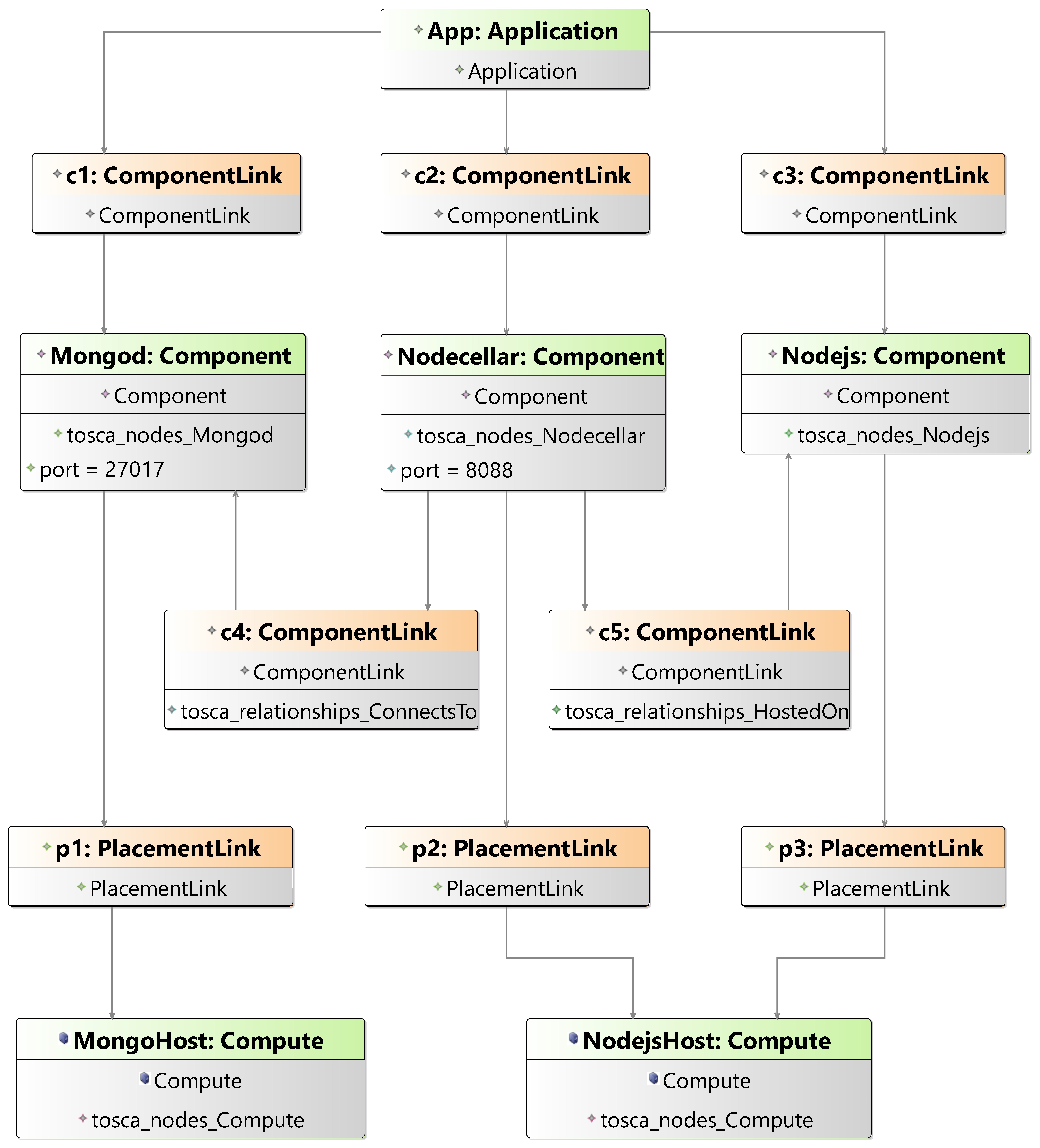}
\caption{Node Cellar Configuration.}
\label{fig:nodecellar}
\end{figure}

The Node Cellar application is a sample JavaScript application that allows to manage (retrieve, create, update, delete) the wines in a wine cellar database. A Node Cellar application can be described using TOSCA types, as depicted in the TOSCA YAML file is depicted of Fig.~\ref{fig:nodecellar-yaml}.	

This topology is automatically transformed into a Node Cellar Configuration using the following mixins defined in TOSCA Extension:

\begin{itemize}
    \item The \textbf{tosca.nodes.Nodecellar} mixin type abstracts the notion of a Node Cellar application and depends on the tosca.nodes.WebApplication mixin type which depends on the MoDMaCAO Component mixin type. It is hosted on a Nodejs server and connected to a MongoDB database.
    \item The {\textbf{tosca.nodes.MongoDB}} mixin type abstracts the notion of a MongoDB database. It depends on the tosca.nodes.DBMS mixin type, which depends on the MoDMaCAO Component mixin type.
    \item The {\textbf{tosca.nodes.Nodejs}} mixin type abstracts the notion of a JavaScript running environment. It depends on the tosca.nodes.WebServer mixin type, which depends on the MoDMaCAO Component mixin type.
    \item The {\textbf{tosca.nodes.Compute}} mixin type abstracts the notion of real or abstract processors of software applications such as \glspl{VM}. It is applied on the Compute resource type.
\end{itemize}

Fig.~\ref{fig:nodecellar} shows the model of a Node Cellar application that corresponds to the topology in Fig.~\ref{fig:nodecellar-yaml}. It is composed of three components (\textsc{nodecellar, nodejs} and \textsc{mongodb})  deployed on two \glspl{VM} (\textsc{NodejsHost} and \textsc{MongoHost}).  OCCI resources and links are represented by boxes in green and orange color, respectively. The application resource is connected to the three component resources via \textsc{ComponentLinks} (\textsc{c1} to \textsc{c3}). The Nodecellar component is connected to the MongoDB component via a \textsc{ConnectsTo} link (\textsc{c4}). The NodeCellar is hosted on the \textsc{Nodejs} component via a \textsc{HostedOn} link (\textsc{c5}). Each component is placed on one \gls{VM} via a \textsc{PlacementLink} (\textsc{p1} to \textsc{p3}). Finally, the properties of all the components and \glspl{VM} are configured. For the sake of brevity, we omit the depiction of \textsc{Attributes} of the components in this model. For illustration, we only keep the attributes regarding the ports used by \textsc{MongoD} and \textsc{Nodecellar}. We can notice that the ports values declared in the YAML file of Fig.~\ref{fig:nodecellar-yaml} have been correctly automatically set in the model. 

\addrevision{
\subsection{Multi-Tier}}

\begin{figure*}
\centering
\includegraphics[width=0.95\linewidth]{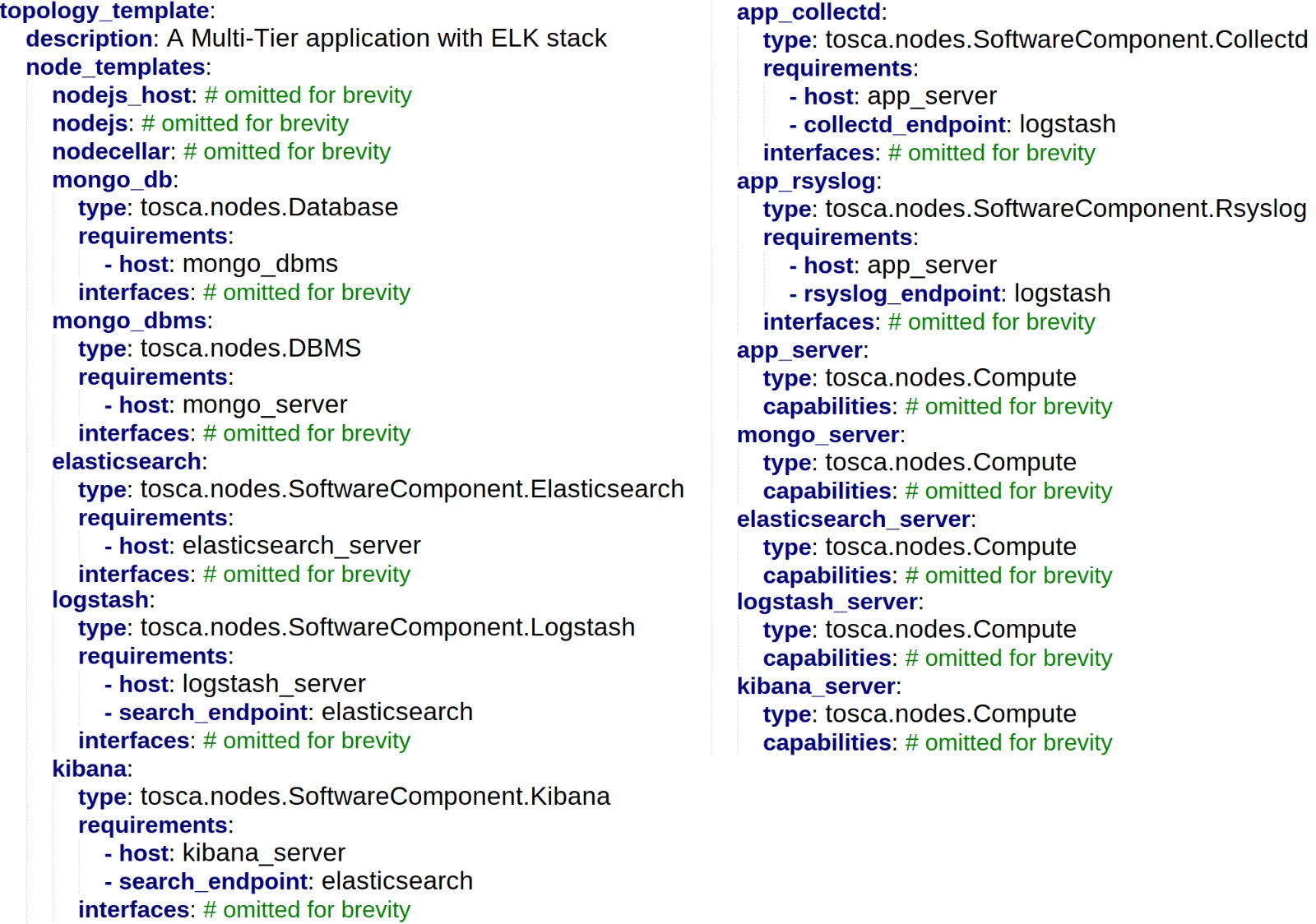}
\caption{Multi-Tier YAML Topology.}
\label{fig:multitier-yaml}
\end{figure*}

\addrevision{
This use case shows the \gls{ELK} stack being used in a typical manner to collect, search and monitor/visualize data from a running application. 
This use case builds upon our NodeCellar application (cf. Section~\ref{sec:nodecellar}) as the one being monitored. 
We successfully describe a Multi-Tier application using TOSCA types, as depicted in the TOSCA YAML file of Fig.~\ref{fig:multitier-yaml}.}

\addrevision{Besides the mixins used to describe a NodeCellar Configuration (cf. Section~\ref{sec:nodecellar}), the Multi-Tier Configuration encompasses the following mixins:}

\addrevision{
\begin{itemize}
    \item The \textbf{tosca.nodes.SoftwareComponent.Elastic\-search} mixin type abstracts the notion of an Elasticsearch search and analytics engine. 
    \item The {\textbf{tosca.nodes.SoftwareComponent.Logstash}} mixin type abstracts the notion of a Logstash data collection engine. 
    \item The {\textbf{tosca.nodes.SoftwareComponent.Kibana}} mixin type abstracts the notion of a Kibana data visualization dashboard. 
    \item The {\textbf{tosca.nodes.SoftwareComponent.Collectd}} mixin type abstracts the notion of a Collectd daemon which collects system and application performance stats. 
    \item The {\textbf{tosca.nodes.SoftwareComponent.Rsyslog}} mixin type abstracts the notion of a an Rsyslog program which transfers log messages over an IP network. 
\end{itemize}
}

\addrevision{All the five mixins defined above depend on the tosca.nodes.SoftwareComponent mixin type which depends on the MoDMaCAO Component mixin type.}

\begin{figure*}
    \centering
    \includegraphics[width=1\linewidth]{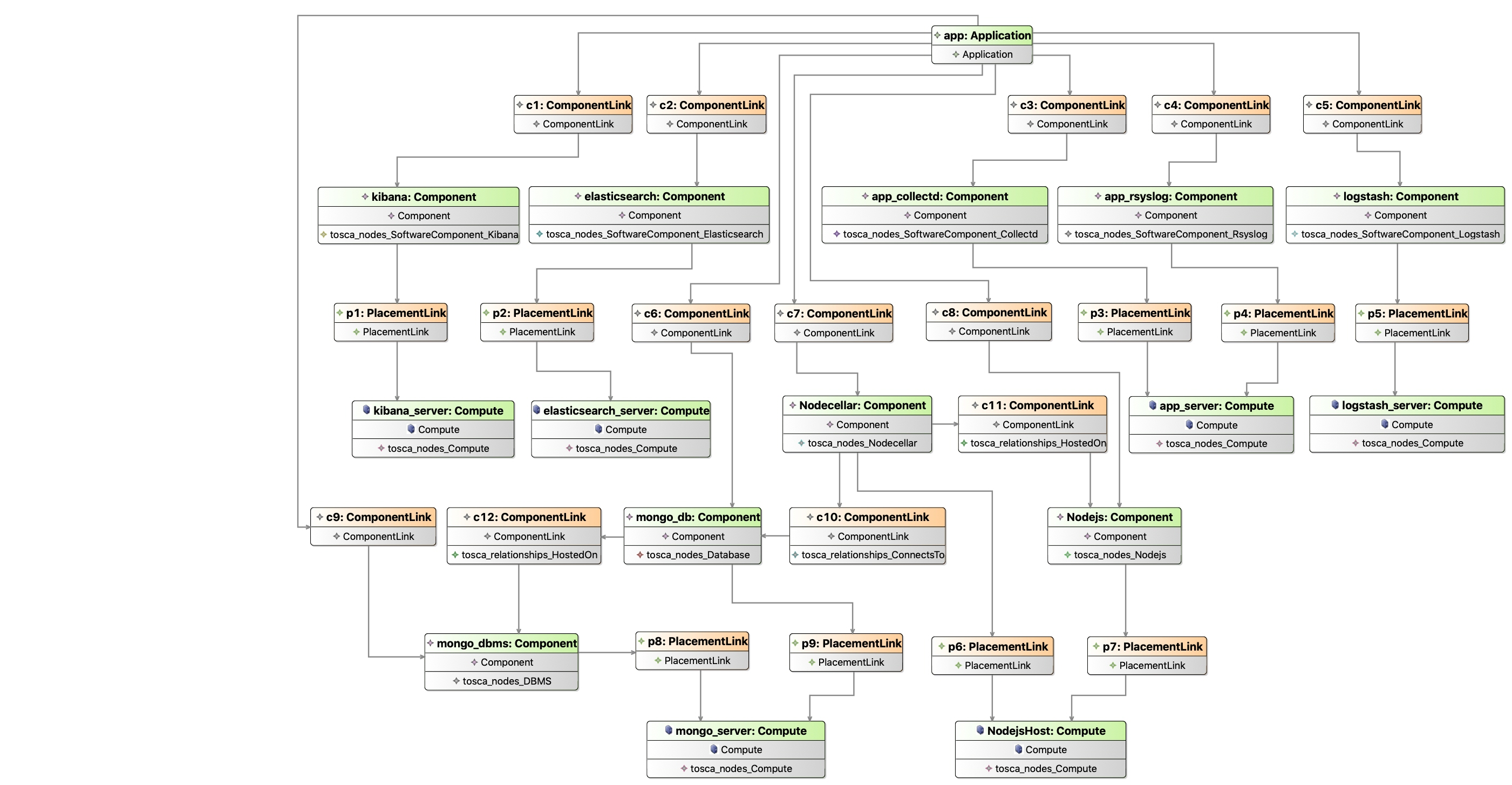}
    \caption{Multi-Tier Configuration.}
    \label{fig:multitier}
\end{figure*}

\addrevision{We show in Fig.~\ref{fig:multitier} the Multi-Tier configuration that corresponds to the YAML topology in Fig.~\ref{fig:multitier-yaml}. It is composed of nine components (\textsc{nodecellar}, \textsc{elasticsearch}, \textsc{logstash}, \textsc{kibana}, \textsc{app\_collectd}, \textsc{app\_\\rsyslog}, \textsc{nodejs}, \textsc{mongo\_db}, \textsc{mongo\_dbms}) deployed on six \glspl{VM} (\textsc{elastic\_server}, \textsc{logstash\_server}, \textsc{ki\\bana\_server}, \textsc{NodejsHost}, \textsc{app\_server}, \textsc{mongo\_\\server}).
Similarly to the two previous use cases, OCCI  resources  and  links  are  represented by boxes in green and orange color, respectively.
The application resource is connected to the nine component resources via \textsc{ComponentLinks} (\textsc{c1} to \textsc{c9}).
Each component is placed on one \gls{VM} via a \textsc{PlacementLink} (\textsc{p1} to \textsc{p9}). \textsc{app\_collectd} and \textsc{app\_rsyslog} share the same VM, as well as \textsc{mongo\_db} and \textsc{mongo\_dbms}.
For brevity, we omit the \textsc{Attributes} of the components in this model.}

\subsection{Orchestration}
\changerevision{Both}{The presented} use case topologies are deployed in the cloud using the model-driven cloud orchestration process. Before the service requests for the individual OCCI resources and links are send to the OCCIware Runtime, the \gls{PIM} to \gls{PSM} transformation is performed on the input configuration, i.e, the WordPress \changerevision{}{, Multi-Tier} or Node Cellar configuration. This transformation ensures that the requirements of the \gls{MoDMaCAO} framework are fulfilled by adding a management network resource to the OCCI configuration. This network ensures, that the \gls{MoDMaCAO} framework has access to each individual Compute node to manage the lifecycle of each modeled component placed on them. Moreover, in case of the WordPress example, this network also connects the Compute nodes \textsc{computeWww} and \textsc{computeDb} to each other, while in the Node Cellar topology the \textsc{MongoHost} and the \textsc{NodejsHost} are linked. \changerevision{}{Also in the Multi-Tier-Deployment the individual Compute nodes are linked to each other.} Thus, in \changerevision{both}{each} use case\changerevision{s}{} the infrastructure required to connect the web server component of (\textsc{WordPress} and \textsc{Nodecellar}) to its database component (\textsc{mysql} and \textsc{Mongod}) is present\changerevision{}{, as well the connection between the \textsc{ELK} components}. It should be noted, that instead of the management network a designated network may be modeled that connects the individual Compute nodes. In addition to the management network, the transformation adds general information to the model, e.g., default SSH keys, user data, flavor and images to be used by the \glspl{VM} to be spawned, which eases the modeling process.

After the transformation, the current cloud deployment is extracted in form of an OCCI model from the OCCIware runtime. Based on the current and desired topology, a provisioning plan is generated~\cite{erbel18closer} that sequences the OCCI requests required to provision and deploy the depicted model. Hereby, the requests are sequenced in such a manner that each Resource is provisioned first, i.e., Compute, Network, Application, and Component. Thereafter, link requests are performed connecting the individual resources with each other. While resources of the platform layer can be immediately linked, Compute nodes have to be in \changerevision{a}{an} active state before they can be connected to networks or storage. These states, amongst others, are reflected in the runtime model and used by the orchestration process. Once the infrastructure has been completely provisioned, i.e., every Compute node being active and connected to the management network, the modeled Applications are deployed. At this point in time, each application including its components are in an undeployed state. Then, depending on the use case, the orchestration process triggers the start action on the WordPress\changerevision{}{, Multi-Tier} or Node Cellar application. This lifecycle management action is implemented by the \gls{MoDMaCAO} framework and triggers the execution of a set of configuration management scripts using the management network provided by the \gls{PIM} to \gls{PSM} transformation. Within these scripts, it is described how to deploy, configure, and start the individual configuration management scripts of the WordPress,\changerevision{ELK, MongoDB}{} and Node Cellar components. In case of the WordPress use case this comprises \textsc{wordpress}, \textsc{php}, \textsc{apache}, and \textsc{mysql}, while in the Node Cellar use case, configuration management scripts describing the management of \textsc{mongod}, \textsc{nodecellar} and \textsc{nodejs} are used. \changerevision{}{Additionally, in the Multi-Tier use case, the components \textsc{elasticsearch}, \textsc{logstash}, and \textsc{kibana} are employed}. Within this deployment process, the installation and execution dependency are respected as modeled within the OCCI model describing the dependencies of the individual components to be deployed, and thus the order in which they get started. A visual documentation of the deployment process of each use case is available in the GitHub repository\footnote{\url{https://github.com/occiware/TOSCA-Studio}}.


%% file: Discussion.tex
\section{Lessons Learned}
Based on our experience with TOSCA and OCCI, we identified two major feedback:
\paragraph{\textbf{Compatibility between TOSCA \& OCCI.}}
Relying a cloud solution on standards is quite advantageous since the latter result of a collective agreement, which means they are accepted in the community, and they also are good in defining the key actions and characteristics of cloud providers.
With the implementation of TOSCA Studio and our two case studies we have successfully demonstrated that both standards can be used orthogonally to implement a model-driven cloud 
orchestration process. We have seen that both provide a similar extension strategy, which can
be exploited to achieve their compatibility. The two standards have a different focus: TOSCA
provides higher level concepts such as the grouping of elements, the definition of policies, and capabilities and requirements, while OCCI provides concepts that mimic runtime behavior, e.g., Mixins that allow to adapt model elements at runtime and a uniform API that allows to create the defined elements on the target cloud infrastructure. TOSCA provides a richer set of modeling elements,
while OCCI is build around a core model which is easier to understand and extend.
In this work, we have successfully demonstrated that the two standards can complement each other, using the strength of TOSCA at design time to model cloud applications and the strengths of
OCCI to actually render API calls from the model to actually provision and deploy the defined cloud resources in a cloud environment.

\paragraph{\textbf{Model-driven design and orchestration of existing cloud applications.}}
Using \gls{MDE} principles, we provided TOSCA Studio, a complete standard-based framework for modeling cloud applications as resources and then concretely provisioning these resources from the cloud. 
For this, we exploited several assets of \gls{MDE} such as \textit{model transformation} when we map TOSCA to OCCI and when we transform the \gls{PIM} to \gls{PSM}, \textit{model verification} when we define structural constraints on TOSCA Extension, \textit{tooling} when we provide TOSCA Studio to have a graphical support of the configurations, and \textit{artifacts generation} when we generate scripts that provision the necessary resources from the cloud. The cherry on the top is the ability of our approach to reuse existing TOSCA topologies and seamlessly ensure their deployment using OCCI API, without any required changes. This does prove the compatibility of our approach with TOSCA and OCCI. 
This framework was successfully tested on \changerevision{two}{three} existing applications WordPress, Node Cellar \addrevision{and Multi-Tier}. We believe it can handle every existing TOSCA topology, even it may require sometimes to enrich TOSCA Extension by adding new TOSCA types.



\label{sec:discussion}

%% file: SoA.tex
\section{Related Work}
\label{sec:stateoftheart}

Besides TOSCA, several other orchestration template formats exist, which have been developed by different cloud providers or communities, e.g., OpenStacks Heat Orchestration Template Language\footnote{\url{https://wiki.openstack.org/wiki/Heat}} and the \changerevision{Amazons}{Amazon's} CloudFormation template format\footnote{\url{https://aws.amazon.com/cloudformation}}. They are not considered in this paper, since our focus is on interoperability of TOSCA and OCCI.
\addrevision{We detail in the following the state-of-the-art of the works around TOSCA and those around OCCI, as well as the works that tackle the integration of standards.}
 
\addrevision{\paragraph{Around TOSCA}}
Andrikopoulos et al. define the GEneralized Topology Language (GENTL)~\cite{andrikopoulos2014gentl} with
the aim to provide a generic modeling language that can easily be mapped
to other concrete, e.g., provider-specific modeling languages that subsequently allow for automated provisioning of the defined resources including TOSCA. They use this language to support the cost-efficient design of application distribution across different cloud provider offerings~\cite{andrikopoulos2014design}.
\changerevision{}{Also Wurster et al.~\cite{wurster2019essential} propose an \gls{EDMM} which is inspired by \gls{TOSCA} to provide a generic language for declarative cloud deployment models.}
Cloudify\footnote{\url{https://cloudify.co/}} is an open source orchestration and management framework for cloud applications lifecycle. It is also based on TOSCA and provides a commercial Web Interface that enables the developer to create deployments and execute workflows. \changerevision{}{Furthermore, web based modeling tools for TOSCA like Winery~\cite{kopp2013winery} exist that visualizes topology models using the Vino4TOSCA language~\cite{breitenbucher2012vino4tosca} that can be provisioned and deployed using OpenTOSCA~\cite{Breitenbuecher2016_OpenTOSCAEcosystem}. Within OpenTOSCA, a uniform interface is defined providing an invocation mechanism for management operations offered by node types~\cite{wettinger2014unified}. While this approach addresses the issue of handling a multitude of proprietary interfaces in TOSCA, OCCI provides a uniform interface by design.} 
\addrevision{Hirmer et al.~\cite{hirmer2014automatic} proposed an approach that completes automatically partial TOSCA topologies in order to make them deployable. The goal is to let the user to be focus on the business-logic and not on technical details. The automatic completion of an uncomplete TOSCA topology is done in two steps: First, it fullfills the requirements.
Then it checks the completness of the topology by trying to automatically provision it. If it is not provisionable, the TOSCA runtime returns a TOSCA topology with missing templates. These missing templates are added to the topology and the first step is repeated, since there are new requirements.}
\addrevision{Brabra et al.~\cite{8814534} propose a model-driven approach based on TOSCA to design resource-related artifacts regardless of specific DevOps tools. The approach enables a new model-driven translation technique that serves to translate the designed artifacts using TOSCA into DevOps specific artifacts and provides connectors that intend to establish the bridge between DevOps-specific artifacts and the DevOps tools.}
\changerevision{GENTL and Cloudify do not consider OCCI that allows models to be executable inside a Models@run.time interpreter framework.}{While a multitude of approaches are based on TOSCA, non of the named approaches consider a connection to OCCI that allows models to be executable inside a Models@run.time interpreter framework.}
With the Eclipse Incubation Project \gls{CAMF}~\cite{loulloudes2015}, Loulloudes et al. attempt to build a whole IDE to manage cloud applications with the help of TOSCA. In the scope of the project different adapters have been developed to deploy the defined TOSCA topology on multiple clouds. However, no model-driven mapping and interaction with OCCI is provided. Regarding the modeling of cloud applications, several extensions to UML have been developed to capture cloud application specifics, e.g.,~\cite{bergmayr2014uml}, \cite{kamali2014ucc}, \cite{guillen2013uml}. In addition, Bergmayr et al.~\cite{bergmayr2016uml2tosca} show how to convert refined UML models to TOSCA templates.
Their approach is also based on an Ecore metamodel generated from the TOSCA XSD. These works consider the modeling of cloud applications, but do not take the mapping to certain API calls into account.

\addrevision{\paragraph{Around OCCI}}
A metamodel for OCCI was defined with help of EMF by~\cite{merle2015precise}, and enhanced by~\cite{zalila2019model}, to provide a common basis for the generation and conformance testing of OCCI tools. This metamodel is used by~\cite{paraiso2016} to model the deployment of applications with help of containers. \addrevision{It is also applied to define a unified metamodel to manage elasticity in the cloud \cite{8742595}.} These works have been published in scope of the OCCIware\footnote{\url{http://occiware.org}} project, that aims to provide a fully integrated IDE to support the whole cloud application management life cycle on multiple clouds based on OCCI. \changerevision{}{Apart from the OCCIware project, approaches exist that utilize OCCI which, e.g, focus on the \gls{PaaS} layer~\cite{Yangui-computer-journal-2016}.} \changerevision{Interoperability}{Still, interoperability} with TOSCA is not considered.

\addrevision{\paragraph{Around standards integration}}
\addrevision{Carrasco et al~\cite{carrasco2014towards} aim at improving the management and the deployment of multi-cloud applications by combining two standards: TOSCA and CAMP. 
Their approach is based on model transformations and allows users to describe their cloud applications according to the distribution of modules and deploy these modules over different clouds.
Their approach is done in two phases: describing the multi-cloud application using Winery and then they transform the TOSCA topology into a CAMP-compliant YAML file in order to deploy the application.}
Later on, Carrasco et al.~\cite{carrasco2016deployment} present an automated approach for the migration of cloud applications components. This approach relies on the trans-cloud framework~\cite{carrasco2018trans}, which is based on the TOSCA topology descriptions and the API of \gls{CAMP} standard~\cite{campSpec}. \addrevision{The main difference between TOSCA-Studio and their approach is the used standards.} Similar to OCCI, CAMP provides a common API for managing cloud providers. However, CAMP targets the deployment of cloud applications on top of PaaS resources, whereas OCCI is suitable for provisioning IaaS, PaaS and SaaS resources. In contrast to~\cite{carrasco2016deployment}, we focus on the deployment and runtime reconfiguration of cloud applications and not on the migration of these applications. Moreover, we propose a resource-based approach, whereas Carrasco et al. propose a component-based approach.




%% file: Conclusions.tex
\section{Conclusion}
\label{sec:conclusions}
Many cloud standards have emerged to cope with the diversity of cloud providers and the heterogeneity encountered in the cloud ecosystem. These standards have different focus work at different levels. 
In this article, we argued that TOSCA and OCCI standards are complementary and we presented an approach to combine TOSCA and OCCI for model and standard driven cloud orchestration. 
We defined an exhaustive and automated mapping between the metamodel elements of TOSCA and OCCI and we adopted this mapping for generating a model for TOSCA that conforms to the OCCIware metamodel (TOSCA Extension). We also proposed TOSCA Studio, a dedicated model-driven environment for designing applications with TOSCA using TOSCA Designer, and for deploying these modeled applications in production environments and adapting them at runtime using OCCI Orchestrator. Furthermore, we used this approach to support the adaptation of models at runtime to keep the model of the infrastructure and the application deployment consistent with its actual state in the cloud. This will also allow us to react to changes in the model or in the cloud.
We also provided \changerevision{two}{three} feasibility studies and showed how WordPress\changerevision{ and}{,} Node Cellar \addrevision{and Multi-Tier} applications can be modeled and concretely deployed using our approach. 

For future work, \addrevision{we plan to support deployment on PaaS by modifying the OCCI orchestrator in order to make requests on a PaaS provider such as Force.com or Cloud Foundry.} We also aim to provide a formal verification of TOSCA Extension by using formal specification languages such as Alloy~\cite{jackson2012software}. Alloy allows to specify TOSCA Extension using first order-logic and to reason about this specification in order to verify desired properties~\cite{challita2018specifying}.
Moreover, by adopting a model-driven approach and automation in our mapping process, it is possible to incorporate changes to both evolving standards and to provide an extensible playground for new concepts. Hence, we aim to extend our catalog of transformation rules by continuously parsing new emerging TOSCA types and adding them to TOSCA Extension. We also aim to support more automated transformation of predefined TOSCA topologies into OCCI configurations. Finally, we plan to conduct a round-trip validation of the deployed application against the designed model, i.e., the configuration.

%% file: paper.bbl
\begin{thebibliography}{10}

\bibitem{occiCore}
Ralf Nyr{\'e}n, Andy Edmonds, Alexander Papaspyrou, Thijs Metsch, and Boris
  Par{\'a}k.
\newblock {Open Cloud Computing Interface - Core}, September 2016.
\newblock [Available online: \url{http://ogf.org/documents/GFD.221.pdf}].

\bibitem{breiter2014software}
Gerd Breiter, Michael Behrendt, M~Gupta, Simon~Daniel Moser, R~Schulze,
  I~Sippli, and Thomas Spatzier.
\newblock {Software Defined Environments based on TOSCA in IBM Cloud
  Implementations}.
\newblock {\em IBM Journal of Research and Development}, 58(2/3):9--1, 2014.

\bibitem{glaser2017tosca2occi}
Fabian Glaser, Johannes~Martin Erbel, and Jens Grabowski.
\newblock {Model Driven Cloud Orchestration by Combining TOSCA and OCCI}.
\newblock In {\em 7th International Conference on Cloud Computing and Services
  Science (CLOSER)}, pages 644--650. SciTePress, 2017.

\bibitem{merle2015precise}
Philippe Merle, Olivier Barais, Jean Parpaillon, No{\"e}l Plouzeau, and Samir
  Tata.
\newblock {A Precise Metamodel for Open Cloud Computing Interface}.
\newblock In {\em {8th IEEE International Conference on Cloud Computing
  (CLOUD)}}, pages 852--859. IEEE, 2015.

\bibitem{zalila2017occiware}
Faiez Zalila, Stéphanie Challita, and Philippe Merle.
\newblock {A Model-Driven Tool Chain for OCCI}.
\newblock In {\em {25th International Conference on COOPERATIVE INFORMATION
  SYSTEMS (CoopIS)}}, pages 389--409. Springer, Cham, 2017.

\bibitem{zalila2019model}
Faiez Zalila, St{\'e}phanie Challita, and Philippe Merle.
\newblock {Model-driven Cloud Resource Management with OCCIware}.
\newblock {\em Future Generation Computer Systems}, 99:260--277, 2019.

\bibitem{rugaber2004model}
Spencer Rugaber and Kurt Stirewalt.
\newblock {Model-Driven Reverse Engineering}.
\newblock {\em IEEE software}, 21(4):45--53, 2004.

\bibitem{blair2009models}
Gordon Blair, Nelly Bencomo, and Robert~B France.
\newblock {Models@ run.time}.
\newblock {\em Computer}, 42(10), 2009.

\bibitem{mdaguide2014}
OMG.
\newblock {MDA Guide rev. 2.0}, 2014.
\newblock OMG Document ormsc/2014-06-01 [Available Online:
  \url{http://www.omg.org/cgi-bin/doc?ormsc/14-06-01.pdf}].

\bibitem{toscaSpec}
OASIS.
\newblock {Topology and Orchestration Specification for Cloud Applications
  (TOSCA) 1.2}, December 2017.
\newblock [Available online:
  \url{http://docs.oasis-open.org/tosca/TOSCA-Simple-Profile-YAML/v1.2/csprd01/TOSCA-Simple-Profile-YAML-v1.2-csprd01.pdf}].

\bibitem{toscaSpecYAML}
OASIS.
\newblock {TOSCA Simple Profile in YAML Version 1.0}, February 2016.
\newblock [Available online:
  \url{http://docs.oasis-open.org/tosca/TOSCA-Simple-Profile-YAML/v1.0/TOSCA-Simple-Profile-YAML-v1.0.html}].

\bibitem{occi-core-12}
Ralf Nyr\'en, Andy Edmonds, Alexander Papaspyrou, Thijs Metsch, and Boris
  Par\'ak.
\newblock {Open Cloud Computing Interface - Core}.
\newblock Specification Document GFD.221, Open Grid Forum, February 2016.

\bibitem{occiInfrastructure}
Thijs Metsch, Andy Edmonds, and Boris Par{\'a}k.
\newblock {Open Cloud Computing Interface - Infrastructure}, September 2016.
\newblock [Available online: \url{http://ogf.org/documents/GFD.224.pdf}].

\bibitem{occiHTTPRendering}
Ralf Nyr{\'e}n, Andy Edmonds, Thijs Metsch, and Boris Par{\'a}k.
\newblock {Open Cloud Computing Interface - HTTP Protocol}, September 2016.
\newblock [Available online: \url{http://ogf.org/documents/GFD.223.pdf}].

\bibitem{breitenbucher2012vino4tosca}
Uwe Breitenb{\"u}cher, Tobias Binz, Oliver Kopp, Frank Leymann, and David
  Schumm.
\newblock {Vino4TOSCA: A Visual Notation for Application Topologies based on
  TOSCA}.
\newblock In {\em OTM Confederated International Conferences" On the Move to
  Meaningful Internet Systems"}, pages 416--424. Springer, 2012.

\bibitem{warmer2003object}
Jos~B Warmer and Anneke~G Kleppe.
\newblock {\em {The Object Constraint Language: getting your models ready for
  MDA}}.
\newblock Addison-Wesley Professional, 2003.

\bibitem{erbel18closer}
Johannes Erbel, Fabian Korte, and Jens Grabowski.
\newblock {Comparison and Runtime Adaptation of Cloud Application Topologies
  based on OCCI}.
\newblock In {\em Proceedings of the 8th International Conference on Cloud
  Computing and Services Science - Volume 1: CLOSER,}, 2018.

\bibitem{paraiso2016}
Fawaz Paraiso, St{\'e}phanie Challita, Yahya Al-Dhuraibi, and Philippe Merle.
\newblock {Model-Driven Management of Docker Containers}.
\newblock In {\em {9th IEEE International Conference on Cloud Computing
  (CLOUD)}}, pages 718--725. IEEE, 2016.

\bibitem{breitenbucher2014declarative}
Uwe Breitenb{\"u}cher, Tobias Binz, Kalman Kepes, Oliver Kopp, Frank Leymann,
  and Johannes Wettinger.
\newblock {Combining Declarative and Imperative Cloud Application Provisioning
  Based on TOSCA.}
\newblock In {\em {IC2E}}, pages 87--96. IEEE Computer Society, 2014.

\bibitem{lushpenko2015adaptation}
Maksym Lushpenko, Nicolas Ferry, Hui Song, Franck Chauvel, and Arnor Solberg.
\newblock {Using Adaptation Plans to Control the Behavior at Runtime}.
\newblock In Nelly Bencomo, Sebastian G{\"o}tz, and Hui Song, editors, {\em
  {CEUR Workshop Proceedings}}, volume 1474. CEUR, 2015.

\bibitem{BreitenbCombining}
U.~{Breitenbücher}, T.~{Binz}, K.~{Képes}, O.~{Kopp}, F.~{Leymann}, and
  J.~{Wettinger}.
\newblock {Combining Declarative and Imperative Cloud Application Provisioning
  Based on TOSCA}.
\newblock In {\em 2014 IEEE International Conference on Cloud Engineering},
  pages 87--96, 2014.

\bibitem{korte2018model}
Fabian Korte, St{\'e}phanie Challita, Faiez Zalila, Philippe Merle, and Jens
  Grabowski.
\newblock {Model-Driven Configuration Management of Cloud Applications with
  OCCI}.
\newblock In {\em 8th International Conference on Cloud Computing and Services
  Science (CLOSER)}, pages 100--111, 2018.

\bibitem{andrikopoulos2014gentl}
Vasilios Andrikopoulos, Anja Reuter, Santiago~G{\'o}mez S{\'a}ez, and Frank
  Leymann.
\newblock {A GENTL Approach for Cloud Application Topologies}.
\newblock In {\em European Conference on Service-Oriented and Cloud Computing},
  pages 148--159. Springer, 2014.

\bibitem{andrikopoulos2014design}
Vasilios Andrikopoulos, Anja Reuter, Mingzhu Xiu, and Frank Leymann.
\newblock {Design Support for Cost-Efficient Application Distribution in the
  Cloud}.
\newblock In {\em 2014 IEEE 7th International Conference on Cloud Computing},
  pages 697--704. IEEE, 2014.

\bibitem{wurster2019essential}
Michael Wurster, Uwe Breitenb{\"u}cher, Michael Falkenthal, Christoph Krieger,
  Frank Leymann, Karoline Saatkamp, and Jacopo Soldani.
\newblock {The Essential Deployment Metamodel: a Systematic Review of
  Deployment Automation Technologies}.
\newblock {\em SICS Software-Intensive Cyber-Physical Systems}, pages 1--13,
  2019.

\bibitem{kopp2013winery}
Oliver Kopp, Tobias Binz, Uwe Breitenb{\"u}cher, and Frank Leymann.
\newblock Winery--a modeling tool for tosca-based cloud applications.
\newblock In {\em International Conference on Service-Oriented Computing},
  pages 700--704. Springer, 2013.

\bibitem{Breitenbuecher2016_OpenTOSCAEcosystem}
Uwe Breitenb{\"u}cherand Christian Endresand~K{\'a}lm{\'a}n K{\'e}pes, Oliver
  Kopp, Frank Leymann, Sebastian Wagner, and Johannes Wettingerand~Michael
  Zimmermann.
\newblock {The OpenTOSCA Ecosystem –Concepts \& Tools}.
\newblock {\em European Space project on Smart Systems, Big Data, Future
  Internet -Towards Serving the Grand Societal Challenges -Volume 1: EPS Rome
  2016}, pages 112--130, 2016.

\bibitem{wettinger2014unified}
Johannes Wettinger, Tobias Binz, Uwe Breitenb{\"u}cher, Oliver Kopp, Frank
  Leymann, and Michael Zimmermann.
\newblock {Unified Invocation of Scripts and Services for Provisioning,
  Deployment, and Management of Cloud Applications Based on TOSCA}.
\newblock In {\em CLOSER}, pages 559--568, 2014.

\bibitem{hirmer2014automatic}
Pascal Hirmer, Uwe Breitenb{\"u}cher, Tobias Binz, Frank Leymann, et~al.
\newblock {Automatic Topology Completion of TOSCA-based Cloud Applications}.
\newblock In {\em GI-Jahrestagung}, pages 247--258, 2014.

\bibitem{8814534}
H.~{Brabra}, A.~{Mtibaa}, W.~{Gaaloul}, B.~{Benatallah}, and F.~{Gargouri}.
\newblock Model-driven orchestration for cloud resources.
\newblock In {\em 2019 IEEE 12th International Conference on Cloud Computing
  (CLOUD)}, pages 422--429, 2019.

\bibitem{loulloudes2015}
N.~Loulloudes, C.~Sofokleous, D.~Trihinas, M.~D. Dikaiakos, and G.~Pallis.
\newblock {Enabling Interoperable Cloud Application Management through an Open
  Source Ecosystem}.
\newblock {\em IEEE Internet Computing}, 19(3):54--59, May 2015.

\bibitem{bergmayr2014uml}
Alexander Bergmayr, Javier Troya, Patrick Neubauer, Manuel Wimmer, and Gerti
  Kappel.
\newblock {UML-based Cloud Application Modeling with Libraries, Profiles, and
  Templates}.
\newblock In {\em {3rd International Workshop on Model-Driven Engineering on
  and for the Cloud (CloudMDE)}}, pages 56--65, 2014.

\bibitem{kamali2014ucc}
Ali Kamali, Soheil Mohammadi, and Ahmad~Abdollahzadeh Barforoush.
\newblock {UCC: UML profile to cloud computing modeling: Using stereotypes and
  tag values}.
\newblock In {\em {7th International Symposium on Telecommunications (IST)}},
  pages 689--694. IEEE, 2014.

\bibitem{guillen2013uml}
Joaqu{\'i}n Guill{\'e}n, Javier Miranda, Juan~Manuel Murillo, and Carlos Canal.
\newblock {A UML Profile for Modeling Multicloud Applications}.
\newblock In {\em {Service-Oriented and Cloud Computing}}, pages 180--187.
  Springer, 2013.

\bibitem{bergmayr2016uml2tosca}
Alexander Bergmayr, Uwe Breitenb{\"u}cher, Oliver Kopp, Manuel Wimmer, Gerti
  Kappel, and Frank Leymann.
\newblock {From Architecture Modeling to Application Provisioning for the Cloud
  by Combining UML and TOSCA}.
\newblock In {\em {6th International Conference on Cloud Computing and Services
  Science (CLOSER)}}, 2016.

\bibitem{8742595}
Y.~{Al-Dhuraibi}, F.~{Zalila}, N.~{Djarallah}, and P.~{Merle}.
\newblock Model-driven elasticity management with occi.
\newblock {\em IEEE Transactions on Cloud Computing}, pages 1--1, 2019.

\bibitem{Yangui-computer-journal-2016}
Sami Yangui and Samir Tata.
\newblock {An OCCI Compliant Model for PaaS Resources Description and
  Provisioning}.
\newblock {\em The Computer Journal}, 59(3):308--324, 2014.

\bibitem{carrasco2014towards}
Jose Carrasco, Javier Cubo, and Ernesto Pimentel.
\newblock {Towards a Flexible Deployment of Multi-Cloud Applications Based on
  TOSCA and CAMP}.
\newblock In {\em European Conference on Service-Oriented and Cloud Computing},
  pages 278--286. Springer, 2014.

\bibitem{carrasco2016deployment}
Jose Carrasco, Javier Cubo, Ernesto Pimentel, and Francisco Dur{\'a}n.
\newblock {Deployment over Heterogeneous Clouds with TOSCA and CAMP}.
\newblock In {\em CLOSER (1)}, pages 170--177, 2016.

\bibitem{carrasco2018trans}
Jose Carrasco, Francisco Dur{\'a}n, and Ernesto Pimentel.
\newblock {Trans-cloud: CAMP/TOSCA-based Bidimensional Cross-Cloud}.
\newblock {\em Computer Standards \& Interfaces}, 58:167--179, 2018.

\bibitem{campSpec}
OASIS.
\newblock {Cloud Application Management for Platforms (CAMP) 1.1}, November
  2014.
\newblock [Available online:
  \url{http://docs.oasis-open.org/camp/camp-spec/v1.1/camp-spec-v1.1.pdf}].

\bibitem{jackson2012software}
Daniel Jackson.
\newblock {\em {Software Abstractions: logic, language, and analysis}}.
\newblock MIT press, 2012.

\bibitem{challita2018specifying}
St{\'e}phanie Challita, Faiez Zalila, and Philippe Merle.
\newblock {Specifying Semantic Interoperability between Heterogeneous Cloud
  Resources with the FCLOUDS Formal Language}.
\newblock In {\em 2018 IEEE 11th International Conference on Cloud Computing
  (CLOUD)}, pages 367--374. IEEE, 2018.

\end{thebibliography}
